\documentclass[a4paper,10pt]{article}
\pdfoutput=1 

\usepackage{jcappub} 

\usepackage[T1]{fontenc} 

\usepackage{jcappub}
\usepackage{amsmath,amssymb,color}
\usepackage{enumerate,xstring}
\usepackage{jcappub}
\usepackage{amsmath}
\usepackage{amsfonts}
\usepackage{amssymb}
\usepackage{graphicx}
\usepackage{epsfig}
\usepackage{color}
\usepackage{multirow}
\usepackage{graphicx}
\usepackage{bigints}
\usepackage{todonotes,bm,etoolbox}
\usepackage{calligra}
\usepackage{hyperref}
\usepackage{tabularx}
\usepackage{booktabs}
\usepackage{soul}
\usepackage{subfigure}
\usepackage{adjustbox}

\usepackage{tikz}

\def\ba#1\ea{\begin{align}#1\end{align}}
\def\bea{\begin{eqnarray}}
\def\eea{\end{eqnarray}}
\def\be{\begin{equation}}
\def\ee{\end{equation}}

\def\({\left(}
\def\){\right)}
\def\[{\left[}
\def\]{\right]}

\def\<{\left\langle}
\def\>{\right\rangle}

\def\comment#1{}

\def\eps{\epsilon}

\renewcommand{\v}[1]{\bm{#1}}

\def\vx{\v{x}}

\def\vq{{\v{q}}}

\def\vr{\v{r}}
\def\vell{\v{\ell}}
\def\vtheta{\v{\theta}}
\def\valpha{\v{\alpha}}

\newcommand{\perm}[1]{ \expandafter\ifstrempty\expandafter{#1} {\mbox{perm.}} {\mbox{$#1$ perm.}} }

\definecolor{RedWine}{rgb}{0.743,0,0}
\definecolor{RoyalBlue}{rgb}{0.25,.41,.88}
\definecolor{ForestGreen}{rgb}{.13,.54,.13}
\definecolor{Goldenrod}{rgb}{.85,.65,.13}

\newcommand{\bq}{\begin{eqnarray}}
\newcommand{\eq}{\end{eqnarray}}

\allowdisplaybreaks

\title{Beyond 3$\times$2-point cosmology: the integrated shear and galaxy 3-point correlation functions}

\author[a,b]{Anik Halder,}
\author[a,b]{Zhengyangguang Gong,}
\author[c,d]{Alexandre Barreira,}
\author[a,c]{Oliver Friedrich,}
\author[a,b]{Stella Seitz,}
\author[a,c]{Daniel Gruen}

\affiliation[a]{Universit\"{a}ts-Sternwarte, Fakult\"{a}t f\"{u}r Physik, Ludwig-Maximilians-Universit\"{a}t M\"{u}nchen,\\Scheinerstra{\ss}e 1, 81679 M\"{u}nchen, Germany}
\affiliation[b]{Max Planck Institute for Extraterrestrial Physics, Giessenbachstra{\ss}e 1, 85748 Garching, Germany}
\affiliation[c]{Excellence Cluster ORIGINS, Boltzmannstra{\ss}e 2, 85748 Garching, Germany}
\affiliation[d]{Ludwig-Maximilians-Universit\"{a}t, Schellingstra{\ss}e 4, 80799 M\"{u}nchen, Germany}

\emailAdd{ahalder@usm.lmu.de}
\emailAdd{lgong@usm.lmu.de}
\emailAdd{alex.barreira@origins-cluster.de}
\emailAdd{oliver.friedrich@physik.uni-muenchen.de}
\emailAdd{stella@usm.lmu.de}
\emailAdd{Daniel.Gruen@lmu.de}

\abstract{We present the integrated 3-point correlation functions (3PCF) involving both the cosmic shear and the galaxy density fields. These are a set of higher-order statistics that describe the modulation of local 2-point correlation functions (2PCF) by large-scale features in the fields, and which are easy to measure from galaxy imaging surveys. Based on previous works on the shear-only integrated 3PCF, we develop the theoretical framework for modelling 5 new statistics involving the galaxy field and its cross-correlations with cosmic shear. Using realistic galaxy and cosmic shear mocks from simulations, we determine the regime of validity of our models based on leading-order standard perturbation theory with an MCMC analysis that recovers unbiased constraints of the amplitude of fluctuations parameter $A_s$ and the linear and quadratic galaxy bias parameters $b_1$ and $b_2$. Using Fisher matrix forecasts for a DES-Y3-like survey, relative to baseline analyses with conventional 3$\times$2PCFs, we find that the addition of the shear-only integrated 3PCF can improve cosmological parameter constraints by $20-40\%$. The subsequent addition of the new statistics introduced in this paper can lead to further improvements of $10-20\%$, even when utilizing only conservatively large scales where the tree-level models are valid. Our results motivate future work on the galaxy and shear integrated 3PCFs, which offer a practical way to extend standard analyses based on 3$\times$2PCFs to systematically probe the non-Gaussian information content of cosmic density fields.}

\begin{document}
\maketitle
\flushbottom

\section{Introduction}
\label{sec:intro} 
Three popular 2-point statistics employed in weak gravitational lensing surveys are: (i) the 2-point cosmic shear correlation function $\xi_{\pm}$, (ii) the angular clustering of foreground lens galaxies $\xi_{g}$, and (iii) the average tangential shear signal of source galaxies around foreground lens galaxies $\xi_t$. Together they are known as the 3$\times$2-point correlation functions (3$\times$2PCFs), jointly probing projections of the late-time power spectrum of matter and galaxy density perturbations. These statistics form a key analysis tool in current surveys such as DES \cite{Abbott_2022}, KiDS  \cite{Heymans_2021}, HSC-SSP \cite{Hamana_2020}, and are expected to continue to provide even tighter constraints on cosmological parameters when upcoming missions like Euclid \cite{2011arXiv1110.3193L}, Vera Rubin's LSST \cite{2012arXiv1211.0310L} and the Nancy Roman Space Telescope \cite{2015arXiv150303757S} go online. However, the late-time matter and galaxy density fields are non-Gaussian distributed \cite{Bernardeau_2002} and thus have information contained in higher-order moments that are not captured by 2-point correlation functions alone. Hence, going beyond 2-point statistics and investigating higher-order correlation functions is of great interest as they can enable even tighter constraints on cosmological parameters. Efforts on this front using cosmic shear or galaxy data include the 3-point cosmic shear correlation functions (3PCF) and third-order aperture mass moments \cite{Takada_2004, Schneider2005, Semboloni2013, Fu2014, Secco_2022, Heydenreich_2022}, galaxy-galaxy-galaxy lensing \cite{Schneider_Watts_2005, Linke_2022}, density-split statistics \cite{Friedrich_2018, gruen_2018, Burger_2020,  burger2022}, the lensing aperture mass and convergence PDF \cite{Barthelemy2020, Boyle_2021, Giblin2023}, third-order convergence moments \cite{Jain_1997, Gatti_2022} and shear peak statistics \cite{Harnois-Deraps2021, zurcher2022, Lanzieri2023forecasting}. These works focus mostly on cosmic shear data, with only a few analysing the galaxy and shear fields together. In particular, a robust framework for joint galaxy and shear cross-correlation analyses for a higher-order equivalent to the 3$\times$2PCFs that can be directly applied to galaxy imaging data to obtain improved cosmological constraints has not been developed so far.

\begin{figure}
\centering 
\includegraphics[width=\textwidth]{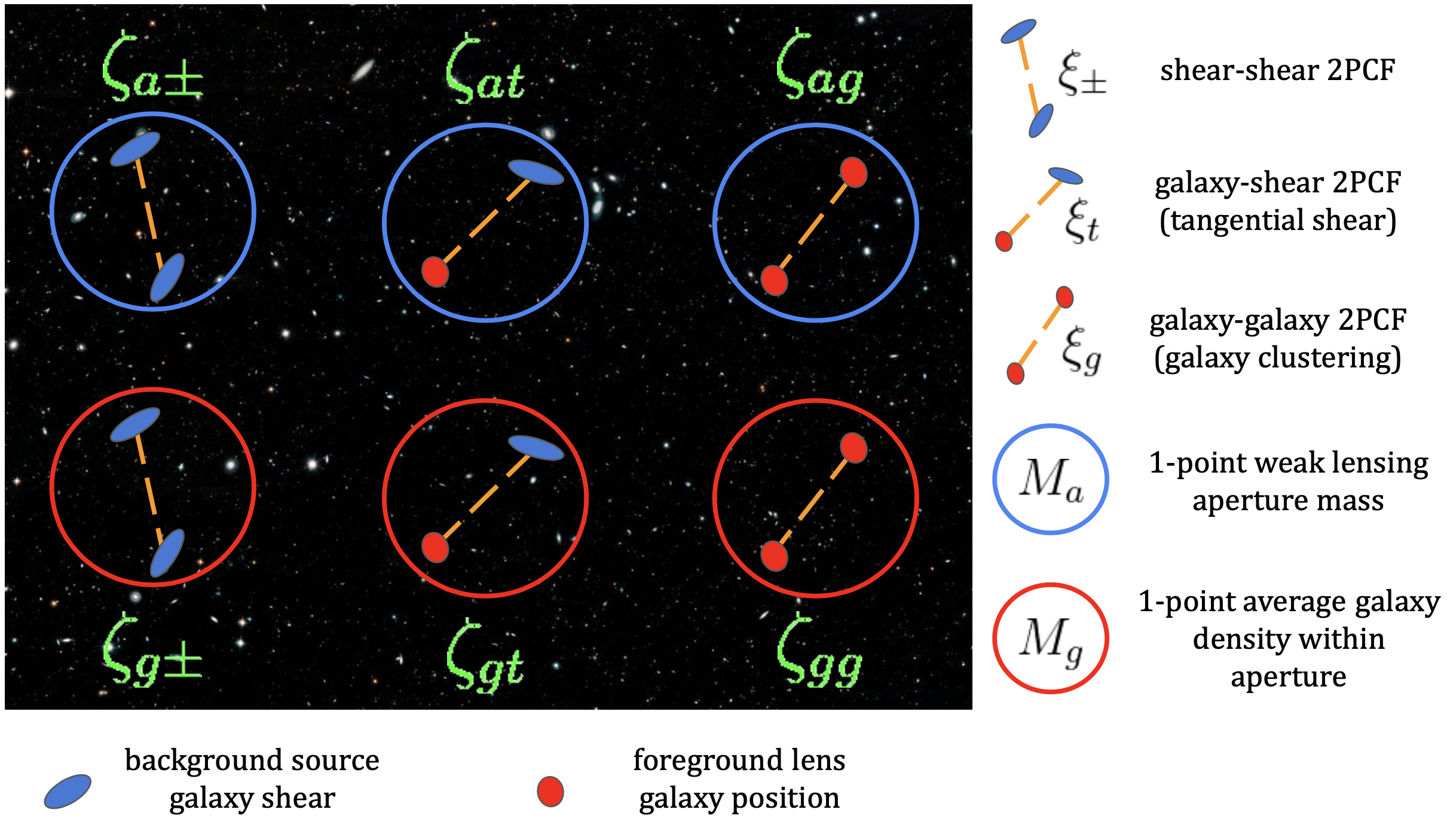}
\caption{\label{fig:6xi3PCFs_illustrations} Illustration showing the ingredients needed for computing the integrated 3PCFs using both cosmic shear (filled blue ellipses) and foreground lens galaxy (filled red ellipses) fields. The ingredients are the position-dependent shear 2PCFs $\xi_{\pm}$, position-dependent tangential shear 2PCFs $\xi_{t}$ and the position-dependent galaxy clustering 2PCFs $\xi_{g}$ measured inside apertures on the sky (left, middle and right, respectively). Also needed are the 1-point weak lensing aperture mass $M_{a}$ (blue circles) and the 1-point average galaxy density contrast $M_{g}$ (red circles), both measured in the same apertures/patches. The 6 integrated 3PCFs $\zeta_{a\pm}, \zeta_{at}, \zeta_{ag}, \zeta_{g\pm}, \zeta_{gt}, \zeta_{gg}$  (left to right starting from top-left in the illustration) are measured by correlating the 3 position-dependent 2PCFs and the 2 aperture masses. Background image: Hubble Legacy Field. Credit: NASA, ESA and Hubble Legacy Field team.}
\end{figure}

Previous works \cite{Halder2021,Halder2022,Gong2023cosmology} have developed a practical higher-order cosmic shear 3-point statistic called the {\it integrated shear 3-point correlation function} $\zeta_{a\pm}$, which measures the correlation of the 1-point lensing aperture mass and position-dependent cosmic shear 2-point correlation function measured within sub-patches of the sky.\footnote{Readers interested in the development history of the integrated 3PCF are referred to Refs.~\cite{Chiang_2014, Chiang_2015, Munshi_2017, jung2021integrated, Halder2021, Halder2022, Gong2023cosmology} for details.} This statistic admits a well defined physical interpretation as the modulation of the small-scale shear 2PCF by long-wavelength features of the cosmic shear field. In this paper, we extend this framework to also include the projected foreground galaxy density field and its cross-correlations with the shear field at the integrated 3PCF level. The new integrated 3PCFs that we introduce in this paper are obtained by (i) measuring a position-dependent shear 2PCF, a position-dependent galaxy 2PCF, and a position-dependent tangential shear signal (i.e.~a galaxy-shear cross-2PCF) within survey patches, and then (ii) correlating each of them with either a 1-point lensing aperture mass $M_a$ or a 1-point average number density of lens galaxies $M_g$ measured at the same patch locations. This yields a total of 6 galaxy and shear integrated 3-point correlation functions $\zeta$, as illustrated in Fig.~\ref{fig:6xi3PCFs_illustrations}.

Concretely, in this paper we introduce these new statistics, derive analytical expressions for them, and evaluate them using either leading-order standard perturbation theory (SPT) or the response function (RF) approach to perturbation theory \cite{Barreira2017}. For the integrated 3PCFs that involve the galaxy density field (which are all except the $\zeta_{a\pm}$ statistic studied previously in Refs.~\cite{Halder2021,Halder2022,Gong2023cosmology} that can be computed accurately in the nonlinear regime using RF), we investigate the regime of validity of our SPT models against measurements from realistic DES-like shear and galaxy mocks. Through Fisher forecasts for a DES-like survey, we also investigate the constraining power of these statistics relative to the conventional 3$\times$2PCFs. We find that the addition of $\zeta_{a\pm}$ (the integrated 3PCF computed from only the shear field) already allows for significant improvements of $20-40\%$ on parameters such as the cold-dark matter density $\Omega_{\rm cdm}$, amplitude of primordial density fluctuations $A_s$, dark energy equation of state parameter $w_0$, reduced Hubble parameter $h$ and even the linear galaxy bias parameter $b_1$. Adding the galaxy integrated 3PCFs and utilizing only the large scales where their leading-order SPT models are valid allows for further improvements of $\sim10\%$, which can increase to $\sim20\%$  when investigating extended cosmologies such as with a dynamical dark energy equation of state parameter. Our results strongly motivate further work on these practical higher-order galaxy and shear integrated 3-point statistics, which have the potential to improve cosmological, as well as galaxy bias parameter constraints using real survey data.

The rest of this paper is as follows. In Secs.~\ref{sec:global2pt}, \ref{sec:posdep} and \ref{sec:zeta} we present the theory expressions for the 3$\times$2PCFs, the position-dependent 1- and 2-point statistics, and the 6 integrated 3PCFs $\zeta$. We present the lensing and galaxy mocks we use in this work in Sec.~\ref{sec:simulations}, and our data vector measurements and covariance estimation in Sec.~\ref{sec:measurements}. In Sec.~\ref{sec:results} we present the comparison of our theoretical models against simulations (Sec.~\ref{sec:comparison_model_sims}), the validation of our models using Monte Carlo Markov Chain (MCMC) analyses (Sec.~\ref{sec:validation_MCMC}), and Fisher forecasts (Sec.~\ref{sec:Fisher}) to investigate the constraining power of the integrated 3PCFs on cosmological and galaxy bias parameters. We summarise and conclude in Sec.~\ref{sec:conclusion}.  Appendix \ref{app:2D_projected_fields} describes more details about the 2D projected shear and galaxy density fields. In App.~\ref{app:HOD} we present expressions for the galaxy bias terms calculated in the halo occupation distribution (HOD) formalism. Appendix \ref{app:3D_spectra} details the expressions of the galaxy-matter power- and bi-spectrum models at leading order in SPT. In App.~\ref{app:point_mass} we discuss details of the point mass contribution to the correlations involving the tangential shear 2PCF.

\section{2-point correlations of shear and galaxy fields}
\label{sec:global2pt}

We study the correlations of two projected cosmic density fields, namely the weak lensing shear field and the projected galaxy density contrast field.\footnote{To simplify the language, we will drop `projected' in `projected galaxy density field' from hereon.} We refer the reader to App.~\ref{app:2D_projected_fields} for further details about these two fields.

The weak lensing convergence $ \kappa(\vtheta)$ can be expressed as a line-of-sight projection of the 3D matter density contrast field $\delta_m^{\mathrm{3D}}$ \cite{BartelmannSchneider2001, Schneider_2006, Kilbinger_2015}
\begin{equation} 
\begin{split}
    \kappa(\vtheta) & = \int_0^{\chi_{\mathrm{lim}}} \mathrm{d} \chi \; q_{\kappa}(\chi) \; \delta_m^{\mathrm{3D}}(\vx, \tau), 
\end{split}
\end{equation}
where $\vx = (\chi\vtheta,\chi)$ is the 3D comoving position with $\chi$ the comoving radial coordinate, $\tau = \tau_0 - \chi$ the conformal time coordinate with $\tau_0$ the present day conformal time, and $q_{\kappa}(\chi)$ the lensing projection kernel for source galaxies which follow a normalized distribution $p(\chi')$:
\begin{equation} 
q_{\kappa}(\chi) = \frac{3H_0^2\Omega_{\rm m}}{2c^2}\frac{\chi}{a(\chi)} \int_\chi ^{\chi_{\rm lim}}{\rm d}\chi' p(\chi')\frac{\chi'-\chi}{\chi'}.
\end{equation}
Here, $\chi_{\rm lim}$ is the upper integral limit of the comoving coordinate usually taken to be the size of the comoving horizon of the observable Universe and $a(\chi)$ is the scale factor of the Universe parametrized in terms of $\chi$ which we also utilise as a time coordinate instead of $\tau$. $H_0$ is the Hubble parameter and $\Omega_{\rm m}$ is the total matter density parameter today. The weak lensing shear $\gamma(\vtheta) = \gamma_1(\vtheta) + i \gamma_2(\vtheta)$ at a given angular position $\vtheta$ on the sky, which can be directly estimated using source galaxy shapes, is a complex quantity where the shear components $\gamma_1$ and $\gamma_2$ are specified in a chosen Cartesian frame (we will work in 2D flat-sky). One can relate this complex shear field to the convergence field through second-order derivatives of the 2D lensing potential. In Fourier space, the shear $\gamma(\vell)$ is related to the convergence $\kappa(\vell)$ via\footnote{To ease the notation, we distinguish between real- and Fourier-space variables by their arguments. For example, $\kappa(\vtheta)$ and $\kappa(\vell)$ are the lensing convergence representations in real and Fourier space, respectively.} 
\begin{equation} \label{eq:shear_convergence_FS_relation}
    \gamma(\vell) \; = \; \frac{(\ell_x + i \; \ell_y)^2}{\ell^2} \; \kappa(\vell) \; = \; e^{2i\phi_{\vell}} \kappa(\vell) \; ;  \qquad \text{for } \ell \neq 0 , 
\end{equation}
where $\ell = \sqrt{\ell_x^2 + \ell_y^2}$ and $\phi_{\vell} = \arctan ( \ell_y / \ell_x )$ is the polar angle of $\vell$.

The local comoving number density of galaxies at comoving position $\vx$ and time $\tau$ can be written in terms of the number density contrast of galaxies $\delta_{g}^{\mathrm{3D}}(\vx,\tau)$ as
\begin{equation} \label{eq:3D_galaxy_number_density}
    n_g^{\mathrm{3D}}(\vx, \tau) = \Bar{n}_g^{\mathrm{3D}}(\tau)[1+\delta_{g}^{\mathrm{3D}}(\vx,\tau)],
\end{equation}
where $\Bar{n}_g^{\mathrm{3D}}(\tau)$ is the cosmic mean comoving number density of galaxies. The 2D projected galaxy density contrast field $\delta_{g}^{\mathrm{2D}}(\vtheta)$ can then be defined as a line-of-sight projection of $\delta_{g}^{\mathrm{3D}}$
\begin{equation}
\begin{split}
    \delta_{g}^{\mathrm{2D}}(\vtheta) = \int \mathrm{d} \chi \; q_g(\chi) \; \delta_{g}^{\mathrm{3D}}(\vx,\tau),
\end{split}
\end{equation}
where $q_g(\chi)$ is a normalized projection kernel of galaxies, which in this case is identified as the observed distribution of foreground lens galaxies i.e.~$q_g(\chi) = p(\chi)$. The $\delta_g^{\mathrm{3D}}$ field is considered to be a \textit{tracer} of the underlying matter field $\delta \equiv \delta_m^{\mathrm{3D}}$, which can be expressed using a series expansion of \textit{operators} $\mathcal{O}$ with accompanying \textit{bias} coefficients $b_{\mathcal{O}}(\tau)$ and \textit{stochasticity} parameters $\epsilon_{\mathcal{O}}(\vx,\tau)$ (see Ref.~\cite{biasreview} for a comprehensive review):
\begin{equation}
\label{eq:bias_expansion}
\delta_{g}^{\mathrm{3D}} (\vx,\tau) = \sum_{\mathcal{O}} b_{\mathcal{O}}(\tau) \mathcal{O}(\vx,\tau) + \Big[ \epsilon(\vx,\tau) + \sum_{\mathcal{O}} \epsilon_{\mathcal{O}}(\vx,\tau)\;  \mathcal{O}(\vx,\tau) \Big].
\end{equation}
Specifically, up to second order in perturbations this equation can be written as
\begin{equation}
\label{eq:bias_expansion_upto_second_order_RS}
    \delta_{g}^{\mathrm{3D}}(\vx,\tau) = b_{\delta}(\tau)\; \delta(\vx,\tau) + b_{\delta^2}(\tau)\; \delta^2(\vx,\tau) + b_{K^2}(\tau)\; K^2(\vx,\tau) + \Big[ \epsilon(\vx,\tau) + \epsilon_{\delta}(\vx,\tau) \; \delta(\vx,\tau) \Big],
\end{equation}
where $K^2 = K_{ij}K^{ij}$ is the square of the 3D tidal field $K_{ij} = \left( \frac{\partial_i \partial_j}{ \nabla^2 } - \frac{ \delta_{ij}}{3} \right) \delta_{m}^{\mathrm{3D}}$. In this equation we have ignored higher-order spatial derivatives of $\delta_m^{\mathrm{3D}}$. The bias parameters $b_{\mathcal{O}}(\tau)$ are interpreted as the response of the local number density of galaxies $ n_g^{\mathrm{3D}}(\vx, \tau)$ to changes in the amplitude of the operators $\mathcal{O}(\vx,\tau)$; they absorb the complicated details of small-scale galaxy formation and evolution \cite{Barreira_2021}. These bias terms are often expressed using another notation $b_1 \equiv b_{\delta}, b_2 \equiv 2 b_{\delta^2}, b_{s^2} \equiv 2 b_{K^2}$ which we adopt throughout. The terms inside the square brackets in Eqs.~\eqref{eq:bias_expansion} and \eqref{eq:bias_expansion_upto_second_order_RS} denote the non-deterministic (stochastic) part of the galaxy-matter relation which arise due to perturbations on small scales in the underlying $\mathcal{O}$ fields. Similar to the deterministic bias relation, the stochastic relation comes with its own free parameters $\epsilon_{\mathcal{O}}$ and an offset term $\epsilon$.\footnote{We assume the stochastic terms to be Poisson random variables (see Sec.~\ref{app:3D_spectra}). Exploring non-Poisson stochasticity in the context of the integrated 3PCF is left to future work.} We shall consider the bias and stochastic terms to be time-independent inside a given galaxy redshift bin.

Using the (cross-) correlations of the cosmic shear and galaxy density contrast fields, we can construct three 2PCFs :
\begin{itemize}
\item  Cosmic shear 2PCFs $\xi_{\pm}$ (shear-shear) defined by correlating the \textit{rotated shear} $\gamma_{\phi_{\valpha}}$ at two angular positions $\vtheta$ and $\vtheta+\valpha$ on the shear field, where $\gamma_{\phi_{\valpha}}$ at each point is computed along the direction $\phi_{\valpha}$ of the separation vector $\valpha$ between the two points (see App.~\ref{app:2D_projected_fields} or Refs.~\cite{Schneider_2002, Jarvis_2004}),
\begin{equation}
\label{eq:xi_pm}
\begin{split}
    \xi_{+}^{ij}(\alpha) &\equiv \langle \gamma_{\phi_{\valpha}}^i(\vtheta) \gamma_{\phi_{\valpha}}^{j*}(\vtheta + \valpha) \rangle = \int \frac{{\rm d} \ell \ \ell}{2 \pi} \mathcal{P}^{ij}_{\kappa}(\ell)J_{0}(\ell \alpha), \\
    \xi_{-}^{ij}(\alpha) &\equiv \langle \gamma_{\phi_{\valpha}}^i(\vtheta) \gamma_{\phi_{\valpha}}^{j}(\vtheta + \valpha) \rangle = \int \frac{{\rm d} \ell \ \ell}{2 \pi} \mathcal{P}^{ij}_{\kappa}(\ell)J_{4}(\ell \alpha);
\end{split}
\end{equation}

\item Angular galaxy clustering 2PCF $\xi_{g}$ (galaxy-galaxy) measured by correlating two points separated by $\alpha$ on the galaxy density contrast field $\delta_g^{{\rm 2D}}$ \cite{Krause2021dark}:
\begin{equation}
\label{eq:xi_g}
\xi^{ij}_{g}(\alpha) \equiv  \langle \delta_g^{{\rm 2D},i}(\vtheta) \delta_g^{{\rm 2D},j}(\vtheta + \valpha) \rangle = \int \frac{{\rm d} \ell \ \ell}{2 \pi} \mathcal{P}^{ij}_{g}(\ell)J_0(\ell \alpha) ;
\end{equation}

\item Tangential shear 2PCF $\xi_{t}$ (galaxy-shear), which is the cross-correlation of the foreground galaxy density field with the rotated shear of a background source galaxy along the direction of the separation vector $\valpha$ joining the foreground lens and the background source galaxy.\footnote{This 2PCF is also known as \textit{galaxy-galaxy lensing} in literature, but we refrain from calling it so to avoid confusions with the galaxy-galaxy clustering 2PCF.} It can be written as (see App.~\ref{app:point_mass} or Ref.~\cite{Krause2021dark}):
\begin{equation}
\label{eq:xi_t}
\xi^{ij}_{t}(\alpha) \equiv \langle \delta_g^{{\rm 2D},i}(\vtheta) \gamma_{\phi_{\valpha}}^j(\vtheta + \valpha) \rangle = \underbrace{\int \frac{{\rm d} \ell \ \ell}{2 \pi} \mathcal{P}^{ij}_{t}(\ell)J_2(\ell \alpha)}_{\xi^{ij,\rm PT}_{t}(\alpha)} \ + \ \frac{\mathcal{M}^i_t}{\alpha^2},
\end{equation}
where we include the contribution from the so-called \textit{point-mass} term, whose amplitude $\mathcal{M}_t$ is a free parameter of the model. These terms are due to the fact that there are small nonlinear scales, which are not well captured by perturbation theory $\xi^{\rm PT}_t$, but which can still contribute to the signal on large scales due to the nonlocal nature of the tangential shear signal (see App.~\ref{app:point_mass} or Ref.~\cite{MacCrann2017} for more details).
\end{itemize}
In the equations above, the superscripts $i$, $j$ denote tomographic bins of the background shear source galaxies or the foreground lens galaxies. We consider only the so-called E-mode shear fields, for which the imaginary parts of $\xi_{\pm}$ and $\xi_{t}$ vanish. In the last equalities in Eqs.~(\ref{eq:xi_pm}), (\ref{eq:xi_g}) and (\ref{eq:xi_t}), we have related the real space 2PCFs to the corresponding lensing/galaxy (cross-) power spectra through inverse harmonic transforms (with $J_{n}$ being the $n$-th order ordinary Bessel function of the first kind). These spectra can in turn be expressed as line-of-sight projections of the 3D matter/galaxy (cross-) power spectra using the Limber approximation \cite{Limber1954, Kaiser1992, Krause2021dark} (see App.~\ref{app:3D_spectra}):
\begin{subequations}
\begin{align}
\mathcal{P}^{ij}_{\kappa}(\ell) &= \int {\rm d\chi} \frac{q_{\kappa}^{i}(\chi) q_{\kappa}^{j}(\chi)}{\chi^2} P^{\rm 3D}_{mm}\left(\frac{\ell}{\chi}, \chi \right), \\
\mathcal{P}^{ij}_{g}(\ell) &= \int {\rm d\chi} \frac{q_{g}^{i}(\chi) q_{g}^{j}(\chi)}{\chi^2} P^{\rm 3D}_{gg}\left(\frac{\ell}{\chi}, \chi \right), \\
\mathcal{P}^{ij}_{t}(\ell) &= \int {\rm d\chi} \frac{q_{g}^{i}(\chi) q_{\kappa}^{j}(\chi)}{\chi^2} P^{\rm 3D}_{gm}\left(\frac{\ell}{\chi}, \chi \right).
\end{align}
\end{subequations}
Here, we have defined the convergence power spectrum in Fourier space as $(2\pi)^2\mathcal{P}^{ij}_{\kappa}(\ell)\delta_D(\vell + \vell') = \langle\kappa^i(\vell)\kappa^j(\vell')\rangle$, the 2D galaxy number density contrast power spectrum as $(2\pi)^2\mathcal{P}^{ij}_{g}(\ell)\delta_D(\vell + \vell') = \langle\delta_g^{{\rm 2D},i}(\vell)\delta_g^{{\rm 2D},j}(\vell')\rangle$, and the convergence-galaxy cross-power spectrum as $(2\pi)^2\mathcal{P}^{ij}_{t}(\ell)\delta_D(\vell + \vell') = \langle\delta_g^{{\rm 2D},i}(\vell)\kappa^j(\vell')\rangle$. We evaluate only the auto-correlations of galaxies in $\xi_{g}$ within the same foreground galaxy redshift bin, i.e.~$i=j$, because the cross-correlation between galaxy density fields $\xi^{ij}_{g}$ in different redshift bins $i \neq j$ is small.\footnote{Effects such as lensing magnification can induce non-zero correlations between galaxies in different redshift bins \cite{Krause2021dark}, but we defer the modelling of these effects to future work.}

We use \verb|HMCODE| \cite{Mead2015} to evaluate the nonlinear matter power spectrum $P^{\rm 3D}_{mm}$ that enters the calculation of $\xi_{\pm}$. To evaluate $P^{\rm 3D}_{gg}$ and $P^{\rm 3D}_{gm}$, which enter the calculation of $\xi_g$ and $\xi_t$, we will rely on standard perturbation theory (SPT). Concretely, we work to leading order (tree-level) and evaluate these 3D spectra as $P_{gg} = b_1^2P_{mm} + P_{\eps\eps}$ and $P_{gm} = b_1P_{mm}$, where $P_{\eps\eps}$ is the power spectrum of the stochastic field $\eps(\vx)$ (see App.~\ref{app:3D_spectra}). Note that owing to our inability to make predictions for galaxy clustering observations on small, nonlinear scales, the galaxy-related statistics will be limited to larger scales compared to $\xi_{\pm}$.

We note already here that 3$\times$2PCFs analyses are able to simultaneously constrain both the $b_1$ and the $A_s$ parameters, where $A_s$ is the amplitude of the primordial scalar power spectrum; at leading-order, $\xi_{\pm} \propto A_s$, $\xi_{t} \propto b_1 A_s$ and $\xi_{g} \propto b_1^2 A_s$. We will see later that when we also consider the integrated 3PCFs involving the galaxy density field, which display different scalings with $b_1$ and $A_s$, they will allow for further breaking of degeneracies and help put tighter constraints.

\section{Position-dependent statistics of shear and galaxy fields} 
\label{sec:posdep}

Having looked at the global 2PCFs, we turn our attention now to position-dependent quantities, i.e.~statistics of the fields within sub-patches of the survey (cf.~Fig.~\ref{fig:6xi3PCFs_illustrations}).

\subsection{Position-dependent 2-point correlation functions}
\label{sec:posdep2pt}
First, we consider the case of the \textit{position-dependent 2PCFs} which we define as 2PCFs measured inside finite patches. Following the mathematical formalism of Ref.~\cite{Halder2021}, we can write the angle-averaged cosmic shear, galaxy clustering and the tangential shear position-dependent 2PCFs as
\begin{subequations}
\begin{align}
\xi^{ij}_{+}(\alpha; \vtheta_C) &= \frac{1}{A_{\rm 2pt}(\alpha)} \int \frac{{\rm d}\phi_{\valpha}}{2 \pi} \int {\rm d}^2\vtheta \ \gamma_{\phi_{\valpha}}^i(\vtheta; \vtheta_C) \gamma_{\phi_{\valpha}}^{j*}(\vtheta + \valpha; \vtheta_C), \label{eq:xi_p_pd} \\
\xi^{ij}_{-}(\alpha; \vtheta_C) &= \frac{1}{A_{\rm 2pt}(\alpha)} \int \frac{{\rm d}\phi_{\valpha}}{2 \pi} \int {\rm d}^2\vtheta \ \gamma_{\phi_{\valpha}}^i(\vtheta; \vtheta_C) \gamma_{\phi_{\valpha}}^{j}(\vtheta + \valpha; \vtheta_C), \label{eq:xi_m_pd} \\
\xi^{ij}_{g}(\alpha; \vtheta_C) &= \frac{1}{A_{\rm 2pt}(\alpha)} \int \frac{{\rm d}\phi_{\valpha}}{2 \pi} \int {\rm d}^2\vtheta \ \delta^{{\rm 2D},i}_g(\vtheta; \vtheta_C) \delta^{{\rm 2D},j}_g(\vtheta + \valpha; \vtheta_C), \label{eq:xi_g_pd} \\ 
\xi^{ij}_{t}(\alpha; \vtheta_C) &= \frac{1}{A_{\rm 2pt}(\alpha)} \int \frac{{\rm d}\phi_{\valpha}}{2 \pi} \int {\rm d}^2\vtheta \ \delta^{{\rm 2D},i}_g(\vtheta; \vtheta_C) \gamma^j_{\phi_{\valpha}}(\vtheta + \valpha;\vtheta_C), \label{eq:xi_t_pd}
\end{align}
\end{subequations}
where the \textit{windowed rotated shear} with respect to direction $\phi_{\valpha}$ at location $\vtheta$ inside a top-hat aperture $W$ centred at $\vtheta_C$ reads
\begin{equation}
\gamma_{\phi_{\valpha}}(\vtheta;\vtheta_C) = \gamma_{\phi_{\valpha}}(\vtheta) W(\vtheta-\vtheta_C),
\end{equation}
and
\begin{equation}
\delta^{{\rm 2D}}_g(\vtheta; \vtheta_C) = \delta^{{\rm 2D}}_g(\vtheta)W(\vtheta - \vtheta_C)
\end{equation}
is the \textit{windowed 2D galaxy density contrast} at location $\vtheta$ inside the same top-hat. The area normalization factor is given by
\begin{equation}
A_{\rm 2pt}(\alpha) = \int \frac{{\rm d}\phi_{\valpha}}{2 \pi} \int {\rm d}^2\vtheta \ W(\vtheta - \vtheta_C) W(\vtheta + \valpha - \vtheta_C) = \int \frac{{\rm d} \ell \ \ell}{2 \pi} \ W(\ell)^2 J_0(\ell\alpha), 
\end{equation}
where in the last equality we have used the Fourier space representation of the window function and used the fact that we adopt only azimuthally symmetric apertures $W$. Specifically, for a top-hat filter of angular radius $\theta_T$, we have:
\begin{equation}
    W(\vell) = W(\ell) = 2\pi \theta_T^2 \frac{J_1(\ell \theta_T)}{\ell \theta_T}.
\end{equation}

In Fourier space, these statistics can be expressed as (using Eq.~\eqref{eq:shear_convergence_FS_relation} and following similar derivation steps as in Refs.~\cite{Halder2021, Halder2022})
\begin{subequations}
\begin{align}
\xi^{ij}_{+}(\alpha; \vtheta_C) &= \frac{1}{A_{\rm 2pt}(\alpha)}  \int_{\phi_{\valpha}} \int_{\vell_1} \int_{\vell_2} \int_{\vq_1} \kappa^{i}(\vell_1) \kappa^{j}(\vell_2) e^{2i(\phi_{\vell_1}-\phi_{\vell_2})} W(\vq_1)W(\vell_{12}-\vq_1) e^{i(\vq_1-\vell_1)\cdot \valpha} e^{i\vell_{12}\cdot\vtheta_C}, \label{eq:xip_expression} \\
\xi^{ij}_{-}(\alpha; \vtheta_C) &= \frac{1}{A_{\rm 2pt}(\alpha)}  \int_{\phi_{\valpha}} \int_{\vell_1} \int_{\vell_2} \int_{\vq_1} \kappa^{i}(\vell_1) \kappa^{j}(\vell_2) e^{2i(\phi_{\vell_1}+\phi_{\vell_2}-4i\phi_{\valpha})} W(\vq_1)W(\vell_{12}-\vq_1) \nonumber \\
& \qquad \ \qquad \qquad \qquad \qquad \qquad  \qquad \qquad \qquad \qquad \qquad \qquad \qquad \times e^{i(\vq_1-\vell_1)\cdot \valpha} e^{i\vell_{12}\cdot\vtheta_C}, \label{eq:xim_expression} \\ 
\xi^{ij}_{g}(\alpha; \vtheta_C) &= \frac{1}{A_{\rm 2pt}(\alpha)}  \int_{\phi_{\valpha}} \int_{\vell_1} \int_{\vell_2} \int_{\vq_1} \delta^{{\rm 2D},i}_g(\vell_1) \delta^{{\rm 2D},j}_g(\vell_2) W(\vq_1)W(\vell_{12}-\vq_1) e^{i(\vq_1-\vell_1)\cdot \valpha} e^{i\vell_{12}\cdot\vtheta_C}, \label{eq:xig_expression} \\ 
\xi^{ij}_{t}(\alpha; \vtheta_C) &= \frac{-1}{A_{\rm 2pt}(\alpha)}  \int_{\phi_{\valpha}} \int_{\vell_1} \int_{\vell_2} \int_{\vq_1} \delta^{{\rm 2D},i}_g(\vell_1) \kappa^{j}(\vell_2) e^{2i(\phi_{\vell_2}-\phi_{\valpha})} W(\vq_1)W(\vell_{12}-\vq_1) e^{i(\vq_1-\vell_1)\cdot \valpha} e^{i\vell_{12}\cdot\vtheta_C}. \label{eq:xit_expression}
\end{align}
\end{subequations}
The $\phi_{\vell_1}$ and $\phi_{\vell_2}$ are polar angles of the wavevectors $\vell_1, \vell_2$, respectively. We also defined the shorthand notations $\int_{\phi_{\valpha}} \equiv \int {\rm d}\phi_{\valpha}/(2\pi)$, $\int_{\vell} \equiv \int {\rm d}^2\vell/(2\pi)^2$ and $\vell_{12...n} \equiv \vell_1 + \vell_2 + ... + \vell_n$. It can be shown that, as expected, the ensemble averages of these position-dependent 2PCFs give the corresponding global 2PCFs.

\subsection{Position-dependent 1-point statistics}
\label{sec:map}
In order to predict the integrated 3PCFs we also need \textit{position-dependent 1-point statistics}. We consider in particular the 1-point lensing aperture mass statistic and the 1-point average galaxy density contrast.

The \textit{1-point lensing aperture mass}  in shear tomographic bin $i$ is defined as \cite{Schneider_2006}
\begin{equation}
\begin{split}
\label{eq:Map_shear}
M_{a}^i(\vtheta_C) = \int {\rm d}^2\vtheta \  \gamma_{t}^i(\vtheta,\phi_{\vtheta-\vtheta_C}) Q(\vtheta-\vtheta_C) = \int {\rm d}^2\vtheta \ \kappa^i(\vtheta) U(\vtheta - \vtheta_C) = \int_{\vell} \kappa^i(\vell) U(\vell) e^{i\vell \cdot \vtheta_C} ,
\end{split}
\end{equation}
where in the last equality we have used the Fourier space representation of the aperture mass. In practice, this can be measured as a weighted mean of the tangential shear field $\gamma_{t} \equiv \Re[\gamma_{\phi_{\vtheta-\vtheta_C}}]$ inside a compensated filter $Q$ centred at location $\vtheta_C$; the Fourier representation in terms of the lensing convergence is useful from a theoretical modelling perspective. These compensated filters by definition satisfy:
\begin{equation}
\int {\rm d}^2\vtheta \ U(\vtheta - \vtheta_C) = \int {\rm d}^2\vtheta \ Q(\vtheta - \vtheta_C) = 0,
\end{equation}
and hence an area normalisation term for this lensing aperture mass is irrelevant. We adopt the following azimuthally-symmetric form for the $U$ and $Q$ filters \cite{Crittenden2002}:
\begin{align}
    \label{eq:crittenden_compensated_filter_U}
    U(\theta) = \frac{1}{2\pi\theta^2_{ap}}\left(1 - \frac{\theta^2}{2\theta^2_{ap}}\right)\text{exp}\left(-\frac{\theta^2}{2\theta^2_{ap}}\right) \qquad ; \qquad
    Q(\theta) = \frac{\theta^2}{4\pi\theta^2_{ap}}\text{exp}\left(-\frac{\theta^2}{2\theta^2_{ap}}\right),
\end{align}
where the aperture scale of the compensated filter is denoted by $\theta_{ap}$. The Fourier space expression for $U$ (which we use for theoretical predictions) is given by
\begin{equation}
    \label{eq:crittenden_compensated_filter_U_fourier}
    U(\ell) = \int \text{d}^2\boldsymbol{\theta} \ U(\theta) e^{-i\boldsymbol{\ell}\cdot\boldsymbol{\theta}} = \frac{\ell^2\theta^2_{ap}}{2}\text{exp}\left(-\frac{\ell^2\theta^2_{ap}}{2}\right).
\end{equation}

Additionally, we also define the \textit{1-point projected average galaxy density} within a top-hat filter $W$ centred at location $\vtheta_C$ and measured in foreground lens tomographic bin $i$ as
\begin{equation}
\begin{split}
\label{eq:Map_galaxy}
M_{g}^i(\vtheta_C) \equiv \frac{1}{A_{W}}\int {\rm d}^2\vtheta \ \delta^{{\rm 2D},i}_g(\vtheta) W(\vtheta - \vtheta_C) = \frac{1}{A_{W}}\int_{\vell} \delta^{{\rm 2D},i}_g(\vell) W(\vell) e^{i\vell \cdot \vtheta_C}.
\end{split}
\end{equation}
The area normalisation term in this case is given by
\begin{equation}
A_{W} =  \int {\rm d}^2\vtheta \ W(\vtheta - \vtheta_C),
\end{equation}
which is simply the area enclosed by the top-hat filter.

\section{Integrated 3-point correlations of shear and galaxy fields}
\label{sec:zeta}

We now have all of the ingredients needed to compute the integrated 3-point correlation functions involving the cosmic shear and the galaxy density contrast fields. In essence, an integrated 3PCF is simply the correlation between (i) a position dependent 1-point weighted mean within a patch of the survey with (ii) the position-dependent 2PCF measured at the same patch location (see Fig.~\ref{fig:6xi3PCFs_illustrations}). In Refs.~\cite{Halder2021, Halder2022, Gong2023cosmology}, the authors studied the case of the integrated shear 3PCF $\zeta_{a\pm}(\alpha)$,\footnote{In Ref.~\cite{Halder2022} the authors denoted the integrated shear 3PCF as $\zeta_{\pm}$. To be consistent with the notations of the other integrated 3PCFs involving the galaxy field that we introduce in this work, we denote the integrated shear 3PCF as $\zeta_{a\pm}$ with the added subscript `$a$' to specify the involvement of the lensing aperture mass $M_a$ in this statistic.} which corresponds to the correlation between the 1-point lensing aperture mass $M_{a}(\vtheta_C)$ with the position-dependent shear 2PCFs $\xi_{\pm}(\alpha;\vtheta_C)$. With the galaxy density contrast field, we can construct 5 additional such cross-correlations, enabling a total of 6 integrated 3PCFs. The derivation steps are similar for all 6 statistics \cite{Halder2021}, which can be written as
\begin{subequations}
\begin{align}
\zeta^{ijk}_{a\pm}(\alpha) & = \Big< M^{i}_{a}(\vtheta_C) \ \xi^{jk}_{\pm}(\alpha; \vtheta_C) \Big> =  \frac{1}{A_{\rm 2pt}(\alpha)} \int \frac{{\rm d}\ell \ \ell}{2\pi} \ \mathcal{B}^{ijk}_{a\pm}(\ell) J_{0/4}(\ell \alpha), \label{eq:zeta_pm_def} \\
\zeta^{ijk}_{g\pm}(\alpha) &= \Big< M^{i}_{g}(\vtheta_C) \ \xi^{jk}_{\pm}(\alpha; \vtheta_C) \Big> =  \frac{1}{A_W A_{\rm 2pt}(\alpha)} \int \frac{{\rm d}\ell \ \ell}{2\pi} \ \mathcal{B}^{ijk}_{g\pm}(\ell) J_{0/4}(\ell \alpha) , \label{eq:zeta_gpm_def} \\
\zeta^{ijk}_{ag}(\alpha) &= \Big< M^{i}_{a}(\vtheta_C) \ \xi^{jk}_{g}(\alpha; \vtheta_C) \Big>  = \frac{1}{A_{\rm 2pt}(\alpha)} \int \frac{{\rm d}\ell \ \ell}{2\pi} \ \mathcal{B}^{ijk}_{ag}(\ell) J_{0}(\ell\alpha), \label{eq:zeta_ag} \\
\zeta^{ijk}_{gg}(\alpha) &= \Big< M^{i}_{g}(\vtheta_C) \ \xi^{jk}_{g}(\alpha; \vtheta_C) \Big> =\frac{1}{A_{W}A_{\rm 2pt}(\alpha)} \int \frac{{\rm d}\ell \ \ell}{2\pi} \ \mathcal{B}^{ijk}_{gg}(\ell) J_{0}(\ell\alpha), \label{eq:zeta_gg} \\
\zeta^{ijk}_{at}(\alpha) &= \Big< M^{i}_{a}(\vtheta_C) \ \xi^{jk}_{t}(\alpha; \vtheta_C) \Big> = \underbrace{\frac{1}{A_{\rm 2pt}(\alpha)} \int \frac{{\rm d}\ell \ \ell}{2\pi} \ \mathcal{B}^{ijk}_{at}(\ell) J_{2}(\ell\alpha)}_{\zeta^{ijk, \rm PT}_{at}(\alpha)} \ + \ \frac{\mathcal{M}^j_{at}}{\alpha^2}\ , \label{eq:zeta_at} \\
\zeta^{ijk}_{gt}(\alpha) &= \Big< M^{i}_{g}(\vtheta_C) \ \xi^{jk}_{t}(\alpha; \vtheta_C) \Big> = \underbrace{\frac{1}{A_W A_{\rm 2pt}(\alpha)} \int \frac{{\rm d}\ell \ \ell}{2\pi} \ \mathcal{B}^{ijk}_{gt}(\ell) J_2(\ell \alpha)}_{\zeta^{ijk, \rm PT}_{gt}(\alpha)} \ + \ \frac{\mathcal{M}^j_{gt}}{\alpha^2} , \label{eq:zeta_gt}
\end{align}
\end{subequations}
where $i$ labels the source or lens tomographic bin inside which we measure either $M_a$ or $M_g$, and $j$ and $k$ denote the tomographic bins used to compute the three position-dependent 2PCFs $\xi_{\pm}, \xi_{g}, \xi_{t}$. The angle brackets denote ensemble average (or in practice, averaging over all patch positions $\vtheta_C$). 

The equations above write the real-space $\zeta$ in terms of their corresponding Fourier-space counterparts called the {\it integrated bispectra $\mathcal{B}(\ell)$}. These integrated bispectra can be expressed in terms of line-of-sight projections of 3D matter and galaxy (cross-) bispectra using the Limber approximation \cite{Buchalter_2000} as (see App.~\ref{app:3D_spectra})\footnote{In order to perform the numerical integrations in the predictions for $\mathcal{B}$ we use the Monte-Carlo Vegas algorithm \cite{LepageVegas}. Moreover, instead of using the inverse Hankel transform integrals directly to convert the $\mathcal{B}(\ell)$ to real space integrated 3PCFs $\zeta(\alpha)$ (and also the $\mathcal{P}(\ell)$ to real space 2PCFs $\xi(\alpha)$), we use expressions with summation over $\ell$ as given in Ref.~\cite{friedrich2021} (see their Eq.~(9)), which are exact in the curved-sky case and more accurate in that they take into account the finite bin widths in which the correlation functions are measured in the data.} 
\begin{subequations}
\begin{align}
\mathcal{B}^{ijk}_{a\pm}(\ell) &= \int {\rm d}\chi \frac{q_{\kappa}^{i}(\chi)q_{\kappa}^{j}(\chi)q_{\kappa}^{k}(\chi)}{\chi^4} \int_{\vell_1}\int_{\vell_2} B^{\rm 3D}_{mmm}\left(\frac{\vell_1}{\chi}, \frac{\vell_2}{\chi}, \frac{-\vell_{12}}{\chi}; \chi\right) e^{2i\left(\phi_{\vell_2} \mp \phi_{-\vell_{12}}\right)} \nonumber \\ 
& \qquad \ \qquad \qquad \qquad \qquad \qquad \qquad \qquad \qquad \qquad \times U(\vell_1)W(\vell_2+\vell)W(-\vell_{12}-\vell) \label{eq:B_sss_pm}, \\
\mathcal{B}^{ijk}_{g\pm}(\ell) &= \int {\rm d}\chi \frac{q_{g}^{i}(\chi)q_{\kappa}^{j}(\chi)q_{\kappa}^{k}(\chi)}{\chi^4} \int_{\vell_1}\int_{\vell_2} B^{\rm 3D}_{gmm}\left(\frac{\vell_1}{\chi}, \frac{\vell_2}{\chi}, \frac{-\vell_{12}}{\chi}; \chi\right) e^{2i\left(\phi_{\vell_2} \mp \phi_{-\vell_{12}}\right)} \nonumber \\
 &\qquad \qquad \qquad \ \qquad \qquad \qquad \qquad \qquad \qquad \qquad \times W(\vell_1)W(\vell_2+\vell)W(-\vell_{12}-\vell) \label{eq:B_gpm}, \\
\mathcal{B}^{ijk}_{ag}(\ell) &= \int {\rm d}\chi \frac{q_{\kappa}^{i}(\chi)q_{g}^{j}(\chi)q_{g}^{k}(\chi)}{\chi^4} \int_{\vell_1}\int_{\vell_2} B^{\rm 3D}_{mgg}\left(\frac{\vell_1}{\chi}, \frac{\vell_2}{\chi}, \frac{-\vell_{12}}{\chi}; \chi\right) U(\vell_1)W(\vell_2+\vell)W(-\vell_{12}-\vell), \label{eq:B_ag} \\
\mathcal{B}^{ijk}_{gg}(\ell) &= \int {\rm d}\chi \frac{q_{g}^{i}(\chi)q_{g}^{j}(\chi)q_{g}^{k}(\chi)}{\chi^4} \int_{\vell_1}\int_{\vell_2} B^{\rm 3D}_{ggg}\left(\frac{\vell_1}{\chi}, \frac{\vell_2}{\chi}, \frac{-\vell_{12}}{\chi}; \chi\right) W(\vell_1)W(\vell_2+\vell)W(-\vell_{12}-\vell)  \label{eq:B_gg}, \\
\mathcal{B}^{ijk}_{at}(\ell) &= \int {\rm d}\chi \frac{q_{\kappa}^{i}(\chi)q_{g}^{j}(\chi)q_{\kappa}^{k}(\chi)}{\chi^4} \int_{\vell_1}\int_{\vell_2} B^{\rm 3D}_{mgm}\left(\frac{\vell_1}{\chi}, \frac{\vell_2}{\chi}, \frac{-\vell_{12}}{\chi}; \chi\right) e^{2i\phi_{-\vell_{12}}} U(\vell_1)W(\vell_2+\vell)W(-\vell_{12}-\vell)  \label{eq:B_at}, \\
\mathcal{B}^{ijk}_{gt}(\ell) &= \int {\rm d}\chi \frac{q_{g}^{i}(\chi)q_{g}^{j}(\chi)q_{\kappa}^{k}(\chi)}{\chi^4} \int_{\vell_1}\int_{\vell_2} B^{\rm 3D}_{ggm}\left(\frac{\vell_1}{\chi}, \frac{\vell_2}{\chi}, \frac{-\vell_{12}}{\chi}; \chi\right) e^{2i\phi_{-\vell_{12}}} W(\vell_1)W(\vell_2+\vell)W(-\vell_{12}-\vell) \label{eq:B_gt}.
\end{align}
\end{subequations}
For the shear-only $\mathcal{B}_{a\pm}$, Ref.~\cite{Halder2022} showed that the 3D nonlinear matter bispectrum $B^{\rm 3D}_{mmm}$ can be modelled accurately using the response approach to perturbation theory \cite{Barreira2017}. This is the calculation we adopt here, which allows to evaluate $\zeta_{a\pm}$ down to nonlinear scales as a function of cosmological and baryonic physics parameters. For the remainder of the integrated bispectra that involve the galaxy density field, we model the corresponding bispectra at leading-order in perturbation theory; we do not display all of these expressions here, but the interested reader can find them in App.~\ref{app:3D_spectra}. In particular, the various integrated 3PCFs display different scalings of the galaxy bias terms.  Concretely, $\zeta_{a\pm} \propto A_s^2$; $\zeta_{g\pm}, \zeta_{at} \propto \{ b_{1}A_s^2, b_{2}A_s^2, b_{s^2}A_s^2 \}$;  $\zeta_{ag}, \zeta_{gt} \propto \{ b_{1}^2A_s^2, b_{1}b_{2}A_s^2, b_{1}b_{s^2}A_s^2, b_{1}A_s/\bar{n} \}$ and $\zeta_{gg} \propto \{ b_{1}^3A_s^2, b_{1}^2b_{2}A_s^2, b_{1}^2b_{s^2}A_s^2, b_{1}^2 A_s/\bar{n} \}$; here, $\bar{n}$ denotes the mean number density of galaxies in a given tomographic bin. The different sensitivity of the 6 integrated 3PCFs to the galaxy bias and cosmological parameters relative to the 3 global 2PCFs\footnote{See Refs.~\cite{Halder2021,Halder2022} for detailed discussions on the different dependence of $\xi_{\pm}$ and $\zeta_{a\pm}$ on cosmological parameters.}  discussed in Sec.~\ref{sec:global2pt} indicates that  joint analyses of these statistics can help to lift parameter degeneracies, leading to tighter parameter constraints overall.

In Eqs.~(\ref{eq:zeta_at}) and (\ref{eq:zeta_gt}) for $\zeta_{at}$ and $\zeta_{gt}$, we note again the presence of point-mass contributions with amplitude $\mathcal{M}_{at}$ and $\mathcal{M}_{gt}$. These statistics involve the position-dependent tangential shear 2PCF, which is why these parameters are introduced due to the nonlocality of the tangential shear signal. The derivation of these point-mass term contributions is shown in App.~\ref{app:point_mass}.

We also note that the correlations involving the 1-point average galaxy density $M_g(\vtheta_C)$ are susceptible to imaging systematics as they directly probe the number density of galaxies within apertures. The presence of a position-dependent systematic effect affecting the observed foreground lens galaxy number count at different locations $\vtheta_C$ on a survey footprint may therefore impact the measurements of the $\zeta_{g\pm}, \zeta_{gt}, \zeta_{gg}$ statistics. We leave the investigation of the impact of such observational systematic effects in the integrated 3PCFs to future work.

\section{Simulations}
\label{sec:simulations}

In this section we present the simulated data we use in order to test our theoretical models of the integrated 3PCFs discussed in the previous section. We use the publicly available cosmological simulation data from Ref.~\cite{Takahashi2017} (hereafter referred to as the T17 simulations).\footnote{The simulation data products are at \url{http://cosmo.phys.hirosaki-u.ac.jp/takahasi/allsky_raytracing/}.} In our work, we use the full-sky lightcone halo catalogues and cosmic shear lensing maps of the simulation suite. The simulation data products were obtained from a gravity-only N-body simulation in a $\Lambda$CDM cosmology with the following parameters : $\Omega_{\rm cdm} = 0.233,\; \Omega_{\rm b} = 0.046,\; \Omega_{\rm m} = \Omega_{\rm cdm} + \Omega_{\rm b} = 0.279,\; \Omega_{\rm \Lambda} = 0.721,\; h = 0.7,\; \sigma_8 = 0.82\; \mathrm{and}\; n_s = 0.97$. The amplitude of the primordial scalar perturbations $A_s$ corresponding to the T17 value of $\sigma_8$ is $A_s = 2.197 \times 10^{-9}$ (we work with $A_s$ in our paper instead of $\sigma_8$). Halos and sub-halos in the simulation were identified using the six-dimensional phase-space friends-of-friends algorithm \verb|ROCKSTAR| \cite{Behroozi2013}. These halo catalogues  were combined in layers of shells to obtain full-sky lightcone halo catalogues. The simulation boxes were also ray traced using the multiple-lens plane ray-tracing algorithm \verb|GRAYTRIX| \cite{Hamana2015, Shirasaki2015} to obtain weak lensing shear maps for several source redshifts between $z = 0.05$ and $z = 5.3$. We utilize 108 realizations of these data products, obtained from multiple realizations of the T17 simulations.

\subsection{Simulated weak lensing shear maps}

\begin{figure}
\centering 
\includegraphics[width=.6\textwidth]{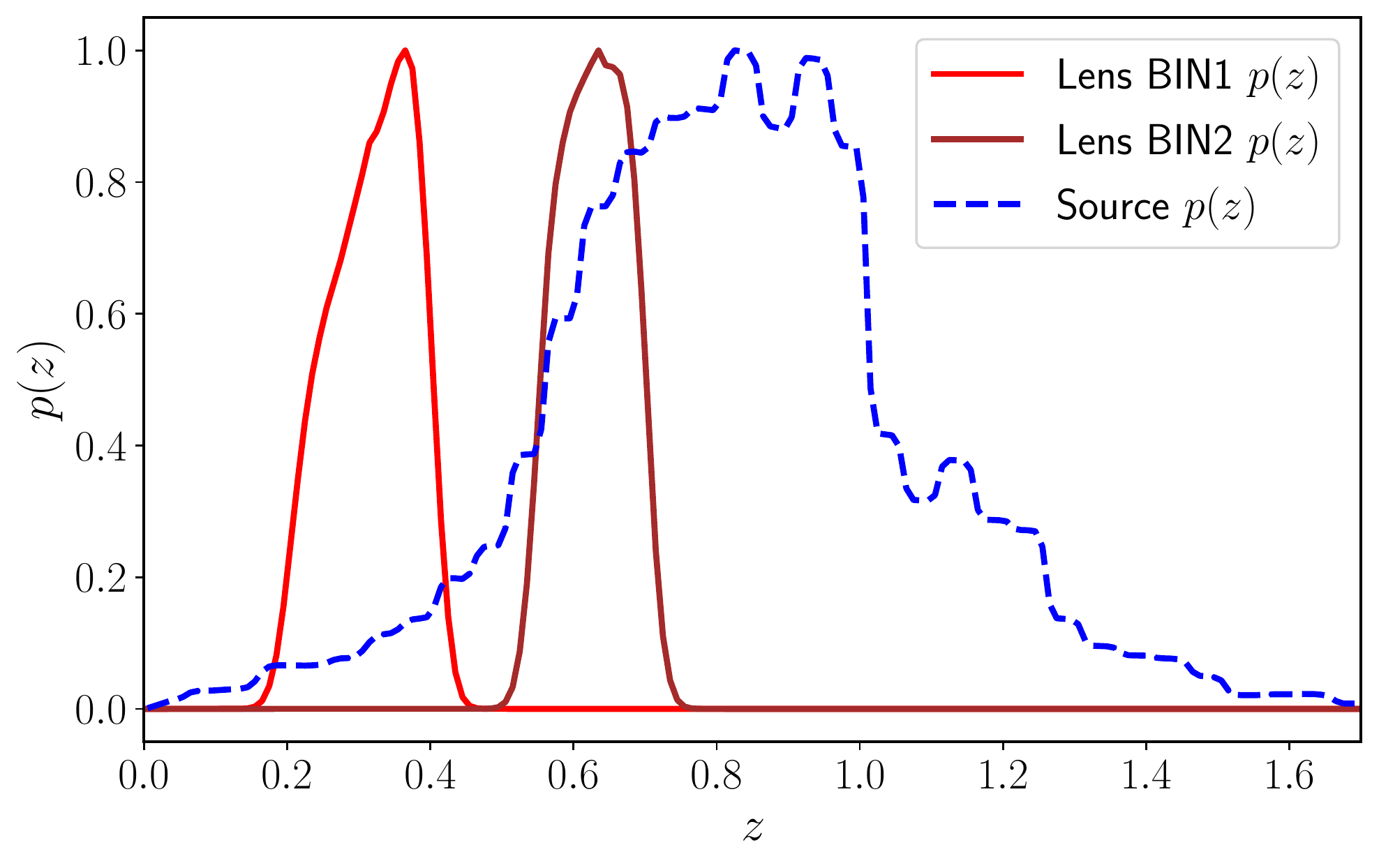}
\caption{\label{fig:nofz} The distribution of lens and source galaxies used for creating our mock galaxy and shear sky maps. The red and brown curves indicate the $p(z)$ of the mock HOD lens galaxies in redshift BIN1 and BIN2, respectively. The source galaxy sample $p(z)$ is shown in blue dashed. For visualization only, the distributions are scaled to have the same maximum at unity.}
\end{figure}

To create a mock shear map from the T17 simulations we use a realistic source galaxy distribution. In order to do that we combine the simulated T17 cosmic shear data products at individual source redshifts according to a source distribution $p(z)$ inspired from the DES Year 3 analysis. We use the same scheme as that used in Ref.~\cite{Gong2023cosmology}, in particular we consider their second tomographic bin (cf.~blue dashed distribution in Fig.~\ref{fig:nofz} of Ref.~\cite{Gong2023cosmology}; this corresponds to a combination of the third and fourth DES Year 3 tomographic bins). The shear map is in \verb|Healpix| \cite{Zonca2019} pixel format to which we add 5 galaxies per ${\rm arcmin}^2$ to mimic the shape noise level expected for DES Year 6 (we refer the reader to Ref.~\cite{Halder2021} for more details about the addition of shape noise).

\subsection{Mock HOD galaxy catalogues}
\label{sec:HOD}

For the purpose of measuring the correlations on the galaxy density field we require mock galaxy catalogues. Being a gravity-only N-body simulation, the T17 suite does not come with galaxy catalogues. We hence create our own full-sky galaxy mocks by populating the T17 halo catalogues using an empirical Halo Occupation Distribution (HOD) method \citep{Berlind_2002} based on the widely used halo model of large-scale structure (see Ref.~\cite{Cooray2002} for a review). Briefly, an HOD model describes a probability distribution $P(N_g | M_h)$ for a halo mass $M_h$ to host $N_g$ galaxies. We follow Ref.~\cite{Zacharegkas_2021} who used a 4-parameter HOD model to investigate the lens galaxy samples used in the DES data. 

Concretely, the HOD model separates the contribution from central and satellite galaxies, and has the following functional forms for the mean values of the central and satellite galaxies, respectively:
\begin{equation}
\label{eq:HOD_mean_N_central}
    \langle N_{cen} \vert M_h \rangle = \frac{1}{2} \Bigg[ 1 + \mathrm{erf} \left( \frac{\log M_h - \log M_{\rm min}}{\sigma_{\log M_h}} \right)\Bigg],
\end{equation}
\begin{equation}
\label{eq:HOD_mean_N_satellite}
    \langle N_{sat} \vert M_h \rangle =  \langle N_{cen} \vert M_h \rangle \left( \frac{M_h}{M_{1}}  \right)^{\gamma}.
\end{equation}
The first equation describes the mean number of central galaxies hosted by halos of mass $M_h$; $M_{\rm min}$ and $\sigma_{\log M_h}$ are the parameters of a smooth step-function. One can understand $M_{\rm min}$ as the mass at which half of the halos with this mass host a central galaxy and $\sigma_{\log M_h}$ gives the scatter around the halo mass $M_h$. The second equation gives the mean number of satellite galaxies within halos of mass $M_h$, and is parametrized by $\gamma$, a power-law index for the mass dependence of the number of satellites and $M_{1}$, the threshold mass for halos to start hosting at least one satellite. The total mean number of galaxies hosted by halos of mass $M_h$ is
\begin{equation}
\label{eq:HOD_mean_N_galaxies}
    \langle N_g \vert M_h \rangle = \langle N_{cen} \vert M_h \rangle +  \langle N_{sat} \vert M_h \rangle.
\end{equation}

Table \ref{tab:HOD_parameters} lists our HOD parameters for two lens tomographic redshift bins that we use in our analyses. They are close to those in Ref.~\cite{Zacharegkas_2021} for their first and third MagLim sample bins (see Table D2 of Ref.~\cite{Zacharegkas_2021}), and as a result, they result in similar galaxy number densities to the MagLim sample. For our HOD parameters, using Eq.~\eqref{eq:galaxy_bias_HOD} in App.~\ref{app:HOD}, we have also computed the galaxy bias parameters $b_{\mathcal{O}}(z)$ averaged over the redshift distributions of the lens bins. We use the halo mass function formula of Ref.~\cite{Tinker_2008} and (i) the halo $b_1$ fitting function from Ref.~\cite{Tinker_2010} to compute $b_1$ for the galaxies, (ii) the halo $b_2(b_1)$ fitting function from Ref.~\cite{Lazeyras_2016} to compute the galaxy $b_2$, and (iii) the co-evolution relation $b_{s^2} = -\frac{4}{7}(b_1 - 1)$ \cite{Baldauf_2012, biasreview} to obtain the galaxy $b_{s^2}$. We take these values to be the fiducial bias parameters of our HOD samples; the bias values are listed in Tab.~\ref{tab:HOD_parameters}.
\begin{table}
\centering
\begin{adjustbox}{width=\textwidth}
\begin{tabular}{|c|c|c|c|c||c||c|c|c|}
\hline
Lens bin & $\log_{10}  M_{\rm min} $ & $\sigma_{\log_{10} M_h}$ & $\log_{10}  M_{1} $ & $\gamma$ & $\Bar{n}_g^{\rm 3D}$ [Mpc$^{-3}$]& $b_1$ & $b_2$ & $b_{s^2}$\\
\hline
BIN1 (z=0.2-0.4) & 12.40 & 0.2 & 13.40 & 0.65 & $1.2 \times 10^{-3}$ & 1.32 &  -0.52 & -0.18\\
BIN2 (z=0.55-0.7)  & 12.03 & $0.014$ &13.37 & 0.52 & $0.4 \times 10^{-3}$ & 1.93 & 0.19 & -0.53\\
\hline
\end{tabular}
\end{adjustbox}
\caption{HOD model parameters used in this work for populating mock lens galaxies in the T17 halo catalogues. Halo masses are expressed in units of $M_{\odot}$.  Also listed are the galaxy number density and bias parameter values obtained for each redshift bin.\label{tab:HOD_parameters}}
\end{table}

To create the actual mock galaxy catalogues from the T17 simulation, we first combine the halo shells to obtain the halos in our two lens redshift intervals. Identifying $M_h$ with $M_{200b}$ (the mass enclosed inside a radius where the mean density is 200 times the background matter density), we use the HOD model described above to populate each halo in the catalogue with galaxies. 
For a given halo we perform a Bernoulli draw with expectation given by Eq.~\eqref{eq:HOD_mean_N_central} to get $N_{cen}$ and a Poisson random draw with expectation given by Eq.~\eqref{eq:HOD_mean_N_satellite} to obtain $N_{sat}$. The central galaxies are placed at the halo centres, whereas the satellite galaxies are placed randomly around the halo centre following a Navarro-Frenk-White distribution \cite{NFW} (this is as in Ref.~\cite{Friedrich_2022}). We note further that we restrict ourselves to using halos with masses $M_{200b} > 1.1\times10^{12} \; M_{\odot}/h$ for BIN1 and $M_{200b} > 5.1\times10^{12} \; M_{\odot}/h$ for BIN2 to remain largely unaffected by the mass resolution limit of the simulation in the respective redshift ranges (see Tab.~1 of Ref.~\cite{Takahashi2017} for mass-cut details).

In order to obtain smoothed looking distribution of lens galaxies in a tomographic bin as expected from photometric galaxy imaging surveys like DES, we first populate HOD galaxies in halos within and beyond the boundaries of the desired redshift range of the tomographic bin. To every true redshift $z_{true}$ of these simulated HOD galaxies, we associate a mock `observed' galaxy redshift $z_{obs} = z_{true} + \delta z$ where $\delta z$ is drawn from a Gaussian with mean 0 and standard deviation $\sigma_z = 0.02$ to mimic a photometric uncertainty in the galaxy's redshift \cite{Rozo_2016,Porredon_2021}. We then select those galaxies whose $z_{obs}$ fall within the tomographic bin's redshift range and use their corresponding $z_{true}$ values to obtain the distribution. The shapes of the mock galaxy distributions in the two redshift bins are shown in Fig.~\ref{fig:nofz}. Finally, we project the galaxy catalogues to 2D grids in \verb|Healpix| format with \verb|NSIDE| $= 2048$  to obtain galaxy number counts maps, which we use to measure the auto- and cross-correlations of the galaxy and cosmic shear fields.

\section{Measurements and data covariance}
\label{sec:measurements}

We measure the galaxy and shear correlations on 6 non-overlapping 5000 ${\rm deg}^2$ circular footprints carved from each all-sky T17 galaxy/shear map; the footprint area is chosen to be representative of a DES-like survey. Over the 108 T17 realizations, this results in a total of 648 DES-sized galaxy and shear maps that we use to obtain our mean data vector and to estimate its covariance.

We follow the same measurement strategy (using the public code \verb|TreeCorr| \cite{Jarvis_2004}) as Ref.~\cite{Halder2021} for the shear-only $\zeta_{a\pm}$ statistic. The position-dependent 2PCFs\footnote{The hat in  $\hat{\xi} (\alpha; \vtheta_C)$ indicates that this is an estimate of the corresponding statistic from data.} $\hat{\xi} (\alpha; \vtheta_C)$ (cf.~Fig.~\ref{fig:6xi3PCFs_illustrations}) are measured on the shear and galaxy density maps within top-hat windows $W$ with radius $\theta_T = 130$ arcmin in 15 log-spaced angular bins within the range $\alpha \in [5,250]$ arcmin. The 1-point lensing aperture mass $\hat{M}_{a} (\vtheta_C)$ at location $\vtheta_C$ is estimated using shear measurements within a compensated window $Q$ with an aperture scale $\theta_{ap} = 130$ arcmin (visualized as blue apertures in Fig.~\ref{fig:6xi3PCFs_illustrations}). The 1-point average galaxy density contrast $\hat{M}_{g}(\vtheta_C)$ on the other hand is measured by taking the mean of the pixel values of the foreground lens galaxy density map within the same top-hat window $W$ where the $\hat{\xi} (\alpha; \vtheta_C)$ are measured (visualized as red apertures in Fig.~\ref{fig:6xi3PCFs_illustrations}). In each 5000 ${\rm deg^2}$ footprint, these apertures are distributed to cover the whole area with only slight overlap between adjacent patches resulting in order 1000 patches across the footprint. The integrated 3PCFs are then estimated as
\begin{equation}
\hat{\zeta}_{xy}(\alpha) = \frac{1}{N_p} \sum_{i=1}^{N_p} \hat{M}_{x}(\vtheta_{C,i}) \ \hat{\xi}_{y}(\alpha, \vtheta_{C,i}) \ ,
\end{equation}
where $x \in \{a, g\}$, $y \in \{\pm, t, g\}$ and the sum runs over the $N_p$ patches centered at $\vtheta_{C,i}$. On the other hand, the global 2PCFs $\hat{\xi} (\alpha)$ in a given footprint are estimated with \verb|TreeCorr| by using all the pixel values for shear or the galaxy density contrast within the entire footprint. 

The full mean data vector is then obtained as the average over the estimates from the 648 mock footprints. For the case of a single lens and single source redshift bin, the data vector consists of the following correlations:
\begin{equation}
\label{eq:data_vector}
d \equiv \{\ \underbrace{\xi_{+}, \xi_{-}, \zeta_{a+}, \zeta_{a-}}_{\text{shear-only}}, \underbrace{\underbrace{\xi_{g}, \zeta_{gg}}_{\text{galaxy-only}}, \underbrace{\xi_{t}, \zeta_{at}, \zeta_{ag}, \zeta_{g+}, \zeta_{g-}, \zeta_{gt}}_{\text{galaxy-shear cross-correlations}}}_{\text{galaxy correlations}} \}.
\end{equation}
To aid in our discussions below, we organize the data vector into different types of contributions. The first 4 components are the \textit{cosmic shear-only} $\xi_{\pm}$ and $\zeta_{a\pm}$ correlations; the 5th and 6th components are the \textit{galaxy-only} correlations $\xi_{g}$, $\zeta_{gg}$; the remaining components correspond to the \textit{galaxy-shear cross-correlations}. We denote the galaxy-only and the galaxy-shear cross-correlations together as \textit{galaxy correlations} indicating that they involve the galaxy field. Accounting for our other lens redshift bin results in additional galaxy correlation terms.

The data covariance matrix is estimated from the mocks as
\begin{equation}
    \label{eq:cov_estimation}
    \mathbf{\hat{C}}= \frac{1}{N_s - 1}\sum_{i=1}^{N_s}(\hat{d}_i - \hat{d})(\hat{d}_i - \hat{d})^{\rm T},
\end{equation}
where $N_s = 648$ is the number of mock footprints, $\hat{d}_i$ is the data vector measured in the $i$-th footprint, and $\hat{d}$ the sample mean over the $N_s$ realizations. To get an unbiased estimate of the inverse covariance matrix we apply the correction \cite{Hartlap2007}
\begin{equation}
    \label{eq:precision_hartlap}
    \mathbf{C}^{-1} = \frac{N_{\rm s} - N_{\rm d} - 2}{N_{\rm s} - 1}\mathbf{\hat{C}}^{-1} \  ,
\end{equation}
where $N_{\rm d} $ is the length of the data vector. For our two lens bins and single source bin we have 20 components in $d$, each with 15 data points making a total of 300 data points before the application of any scale cuts. All of the components of our data vector are shown by the black points with error bars in Fig.~\ref{fig:correlations}; the error bars shown are the square root of the diagonal of the covariance matrix.

\section{Results}
\label{sec:results}

In this section we present our main numerical results for the modelling and cosmological constraining power of the integrated 3PCFs. First, we discuss the regime of validity of our perturbation theory model for the galaxy correlations, i.e.~$\xi_{t}, \xi_{g}, \zeta_{at}, \zeta_{ag}, \zeta_{g\pm}, \zeta_{gt}$. Then, we demonstrate the ability of our galaxy correlation models to recover unbiased constraints on $A_s$, as well as on the bias parameters $b_1$ and $b_2$ of the mock galaxy samples through an MCMC likelihood analysis. Finally, we present Fisher forecast results where we investigate the constraining power of a joint 3$\times$2PCF and integrated 3PCF analysis using all the correlations i.e., both cosmic shear-only ($\xi_{\pm}$, $\zeta_{a\pm}$) as well as galaxy correlations.

\subsection{Comparison of theoretical models to measurements from simulations}
\label{sec:comparison_model_sims}

\begin{figure}
\centering 
\includegraphics[width=\textwidth]{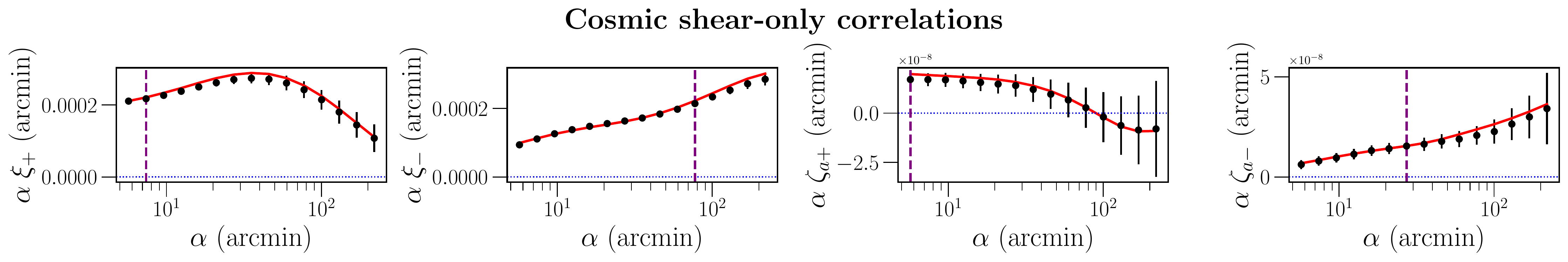}\\
\vspace{0.2cm}
\includegraphics[width=\textwidth]{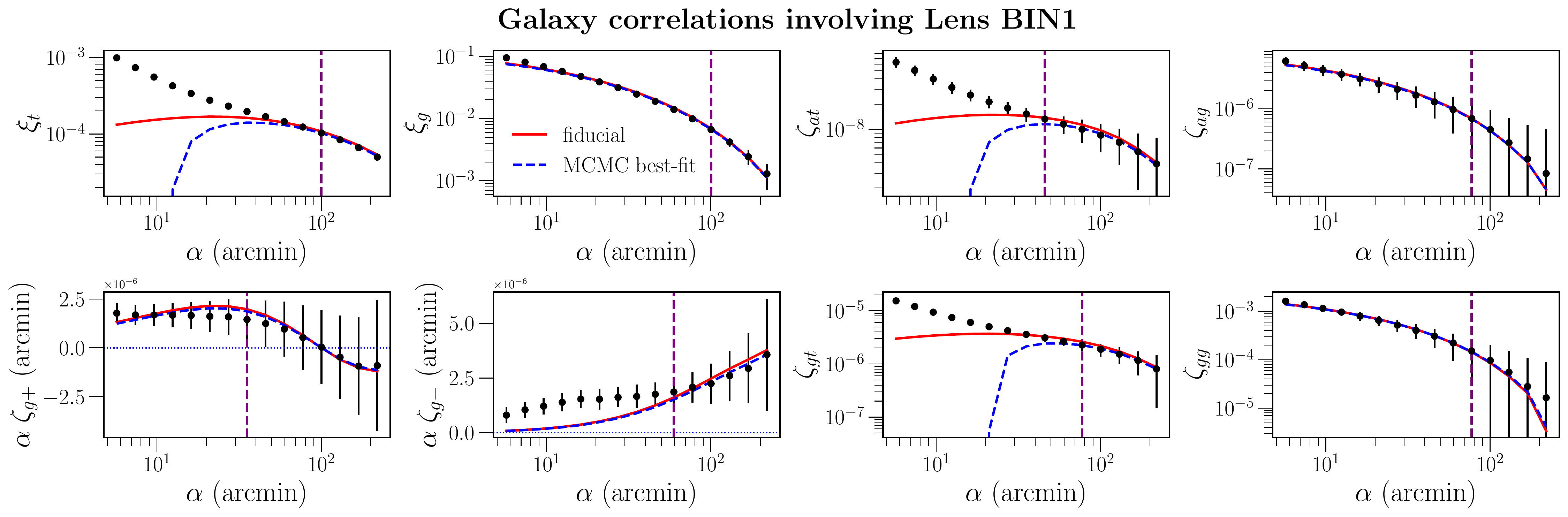}\\
\vspace{0.2cm}
\includegraphics[width=\textwidth]{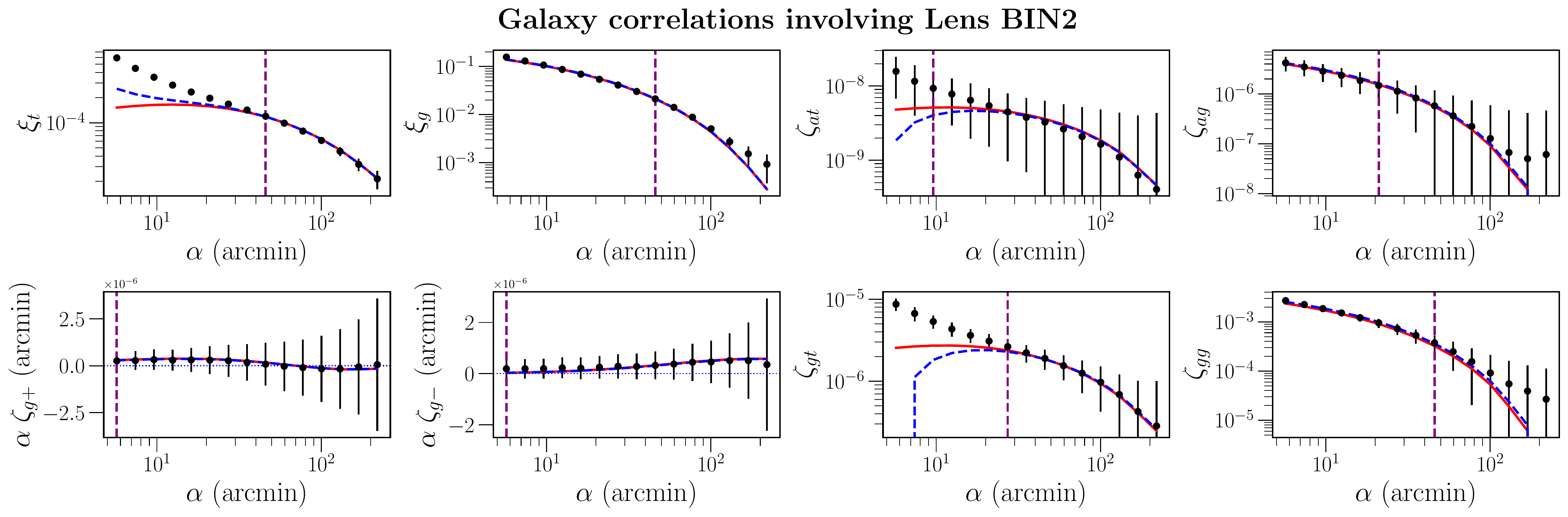}
\caption{\label{fig:correlations} \textbf{First row}: The shear-only 2PCF $\xi_{\pm}$ and the integrated 3PCF $\zeta_{a\pm}$ computed from the T17 mock source tomographic bin. The black dots with the error bars show the mean and the standard deviation of the measurements from our 648 T17 mocks, respectively. The red curves show the model predictions using the fully nonlinear theoretical recipes for these shear correlations computed at the fiducial cosmological parameters of the simulations. The purple dashed vertical lines show our scale cuts to remove scales affected by baryonic feedback effects. \textbf{Second and third rows}: The galaxy 2PCFs $\xi_{t}, \xi_{g}$ and the integrated galaxy 3PCFs $\zeta_{at}, \zeta_{ag}, \zeta_{g\pm}, \zeta_{gt}, \zeta_{gg}$ involving galaxy correlations and cross-shear correlations computed with galaxies in Lens BIN1. The red curves show the tree-level perturbation theory models computed using the fiducial cosmological parameters of the simulations and the fiducial galaxy bias parameter values evaluated using the HOD approach for the lens galaxy sample. The purple dashed vertical lines show our conservative scale cuts to remove scales where our tree-level perturbation theory model breaks down; scales below these cuts are not included in the MCMC and Fisher forecast analyses. The blue dashed curves show the theory predictions computed at the maximum posterior of the MCMC analysis performed in Section~\ref{sec:validation_MCMC}. \textbf{Fourth and fifth rows}: Same as the second and third rows but for Lens BIN2 instead of BIN1.}
\end{figure}

Figure \ref{fig:correlations} compares the components of our data vector in Eq.~\eqref{eq:data_vector} (black dots) with our theory model predictions from Secs.~\ref{sec:global2pt} and \ref{sec:zeta} evaluated at the fiducial cosmology and bias parameters (red curves).

The first row shows the cosmic shear-only correlations, namely the global shear 2PCFs $\xi_{\pm}$ and the integrated shear-only 3PCFs $\zeta_{a\pm}$; recall these are evaluated using the nonlinear matter power spectrum from the \verb|HMCODE| and the {\it response approach} bispectrum model from Ref.~\cite{Halder2022}, respectively. As previously discussed in Ref.~\cite{Halder2022}, the theoretical models for these shear-only correlations are in excellent agreement with the simulation measurements on all angular scales probed. Further, our model allows to readily incorporate the impact of baryonic feedback on small scales. We follow the strategy of Ref.~\cite{Halder2022} (see their Sec.~4.2) to determine scale cuts to remove scales that are expected to be severely affected by baryonic physics. In Fig.~\ref{fig:correlations}, these are shown by the purple dashed vertical lines in the first row of panels; concretely, the data to the right of these lines are deemed to be unaffected by baryonic feedback.

The second and third row of panels in Fig.~\ref{fig:correlations} show the galaxy correlations for Lens BIN1, i.e.~the $\xi_{t}, \xi_{g}$ 2PCFs and the $\zeta_{at}, \zeta_{ag}, \zeta_{g\pm}, \zeta_{gt}, \zeta_{gg}$ integrated 3PCFs. The first and second panels in the second row show $\xi_{t}(\alpha)$ and $\xi_{g}(\alpha)$. For $\xi_{t}$, our tree-level perturbation theory model is in good agreement with the simulation results on only large angular scales $\alpha$, whereas for $\xi_{g}$, the tree-level model displays a good fit down to comparatively smaller angular scales. This is as expected since at a given $\alpha$, $\xi_{t}$ is more sensitive to larger multipoles $\ell$ (smaller nonlinear scales) compared to $\xi_{g}$; this is because the former's $J_2$ Bessel function in the Fourier- to real-space conversion (cf. Eqs.~\eqref{eq:xi_t}, \eqref{eq:xi_g}) weights the nonlinear scales more than the $J_0$ function of the latter. Thus, $\xi_{t}$ gets more contributions from scales where our perturbation theory model breaks down, hence the poorer agreement between theory and simulations in the figure.

The third and fourth panels in the second row show the integrated 3PCFs $\zeta_{at}(\alpha)$ and $\zeta_{ag}(\alpha)$, whereas the panels in the third row are for the integrated 3PCFs $\zeta_{g+}, \zeta_{g-}, \zeta_{gt}, \zeta_{gg}$. The galaxy correlations $\zeta$ show similar trends as $\xi_{t}$, $\xi_{g}$: the models agree with the simulations on large angular scales, but become discrepant on smaller scales where the tree-level models break down (this can be similarly understood in terms of the $J_n$ weightings of each statistic). We note that in our fiducial predictions for $\xi_{t}, \zeta_{at}, \zeta_{gt}$ we set the corresponding point-mass terms $\mathcal{M}$ to zero. The values of these parameters cannot be predicted from first principles (they capture a complicated interplay of higher-order galaxy bias, stochastic terms and nonlinear matter fluctuations), and so in our MCMC validation analysis below we will treat them as free parameters.

We conservatively estimate the regime of validity of our perturbation theory models for the galaxy correlations as follows. We compute the theoretical predictions for two scenarios: (i) using the fiducial tree-level model $d_{\rm tree}$ and (ii) another model $d_{k_{\rm NL}}$ where we artificially set all Fourier modes $k$ larger than the non-linear scale $k_{\rm NL}$ to zero.\footnote{The nonlinear scale is defined implicitly as $k_{\rm NL}^3 P^{\mathrm{3D}}_{\mathrm{lin}}(k_{\rm NL},z) / (2\pi)^3 = 1$.} Using these predictions, we progressively discard small angular scales from our data vector until $\chi^2 \equiv (d_{\rm tree} - d_{k_{\rm NL}})^{\rm T} \mathbf{C}^{-1} (d_{\rm tree} - d_{k_{\rm NL}}) < 0.3$ is satisfied. This roughly identifies a minimum scale $\alpha_{\rm min}$ below which our perturbation theory model begins to fail significantly; the resulting scale cuts are marked by the purple dashed vertical lines in the galaxy correlations panels.\footnote{Note these scale cuts are not the same as those applied on the shear-only statistics, which ensure instead that the scales are not affected by baryonic feedback effects. We assume also that perturbation theory breaks down on scales larger than the scales where baryonic effects are important, as is reasonable.} We note that our criteria for determining these scale cuts are conservative. In particular, allowing the point-mass terms to vary, which we currently set to zero, could allow for greater reach down to smaller angular scales in $\xi_{t}$, $\zeta_{at}$, and $\zeta_{gt}$. This could in turn enable the use of higher signal-to-noise data points. As our primary aim is to explore the first-order information gain that the galaxy correlations can already bring from scales where leading-order PT models are valid, we choose to adopt these conservative scale cuts and defer the investigation of more accurate models on smaller angular scales to future works.

Finally, similar considerations hold for the fourth and fifth rows of panels in Fig.~\ref{fig:correlations}, which show the same as the second and third rows, but for the Lens BIN2. We note only that the error bars in some of the panels are larger than those for BIN1, which is as expected by the fact that the number density of galaxies in BIN2 is approximately $3$ times smaller than that of BIN 1 (cf.~Tab.~\ref{tab:HOD_parameters}).

\subsection{MCMC validation of the galaxy correlations modelling}
\label{sec:validation_MCMC}

\begin{figure}
\centering 
\includegraphics[width=0.7\textwidth]{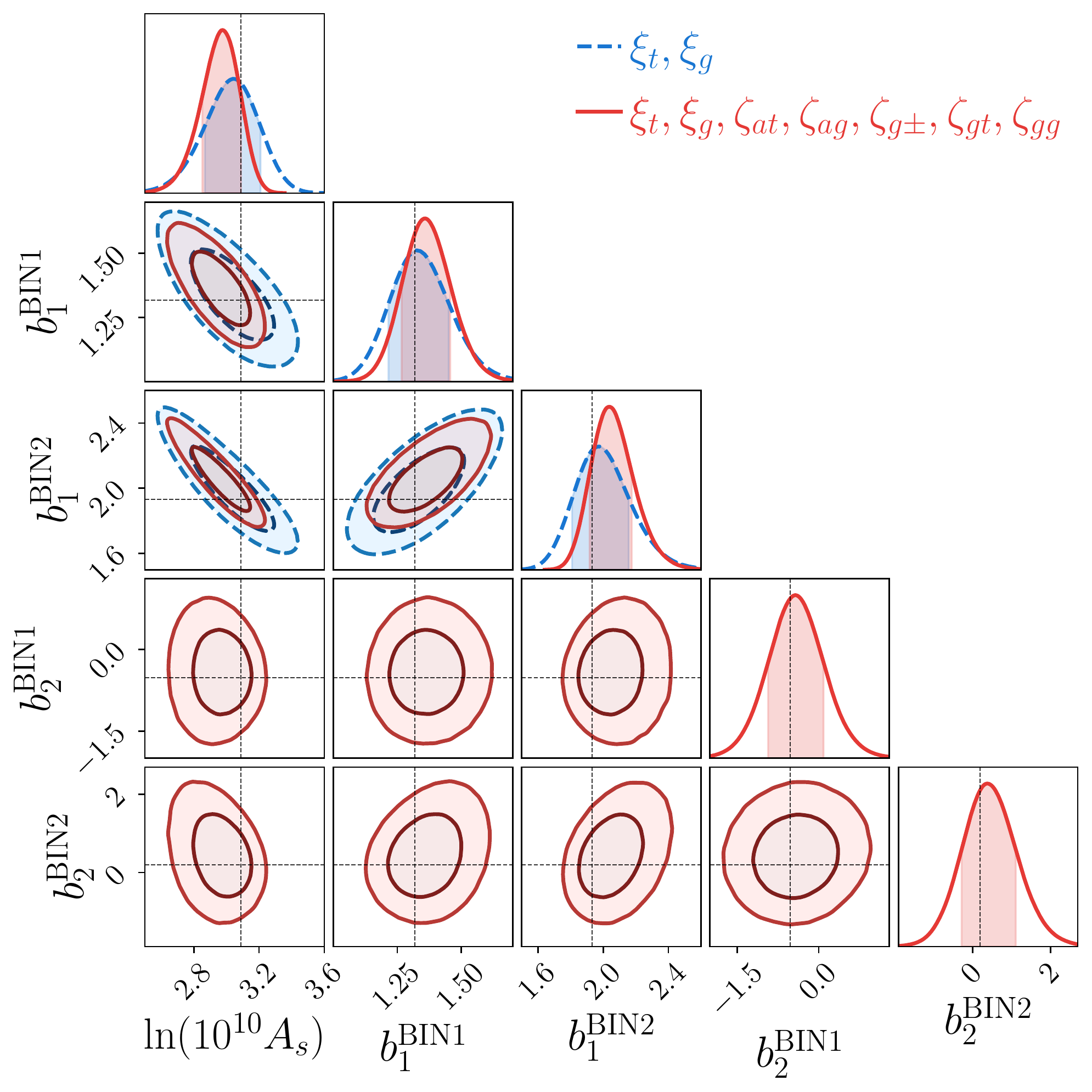}
\caption{\label{fig:MCMC_galaxy_x_shear_marginalised_PM} Marginalized one- and two-dimensional MCMC constraints on the parameters $\ln(10^{10}A_s)$, $b_1^{\text{\tiny{BIN1}}}$, $b_1^{\text{\tiny{BIN2}}}$, $b_2^{\text{\tiny{BIN1}}}$, $b_2^{\text{\tiny{BIN2}}}$ obtained from the galaxy 2PCFs $\xi_{t}, \xi_{g}$ (blue dashed) and considering also the galaxy integrated 3PCFs, $\xi_{t}, \xi_{g}, \zeta_{at}, \zeta_{ag}, \zeta_{g\pm}, \zeta_{gt}, \zeta_{gg}$  (red solid). We consider the average T17 measurements as the data vector with the scale cuts discussed in Sec.~\ref{sec:comparison_model_sims}. The fiducial values of the parameters are marked by the black dashed lines. Note that at tree-level, the galaxy 2PCFs cannot constrain the second-order $b_2$ parameters. The point-mass terms associated with tangential shear correlations are marginalized over; see Fig.~\ref{fig:MCMC_galaxy_x_shear} in App.~\ref{app:point_mass} for their constraints. All other parameters are fixed to the fiducial values of the simulation and the tidal bias terms are varied according to the co-evolution $b_{s^2}(b_1)$ relation.}
\end{figure}

Using the measured galaxy correlations with the scale cuts, we now test whether our tree-level perturbation theory models can correctly recover the fiducial parameters in an MCMC constraint analysis; we do not consider the shear-only correlations in this section to focus on the galaxy correlations modelling. We investigate in particular the constraints on the galaxy bias parameters $b_1$ and $b_2$, as well as the cosmological parameter $A_s$.\footnote{We take $A_s$ as the only cosmological parameter in the MCMC results as predictions for it are rapid to obtain. The extension to other cosmological parameters would require the construction of dedicated emulators for fast theory predictions (e.g.~see Ref.~\cite{Gong2023cosmology} for an emulator for $\zeta_{a\pm}$), which is beyond the scope of this work.} We assume a Gaussian likelihood function and wide uniform priors for the parameters varied: in addition to $A_s$ and the $b_1$, $b_2$ of the two samples, we also vary the point-mass $\mathcal{M}$ amplitude parameters that contribute to the $\xi_{t}, \zeta_{at}, \zeta_{gt}$ statistics. We do not sample $b_{s^2}$ but let it vary according to the co-evolution relation for $b_{s^2}(b_1)$. We use the publicly available affine sampler \verb|emcee| \cite{Foreman_Mackey_2013} to perform the MCMC analysis. The results are shown in the contour plot in Fig.~\ref{fig:MCMC_galaxy_x_shear_marginalised_PM}. The figure is for the result marginalized over the point-mass terms, but see Fig.~\ref{fig:MCMC_galaxy_x_shear} in App.~\ref{app:point_mass} for the constraints on all varied parameters.

The figure shows that our tree-level models for $\xi_{t}$, $\xi_{g}$, which are only sensitive to $A_s$ and $b_1$, correctly recover the fiducial values of the parameters within the 68\% credible intervals (blue dashed contours). When considering the galaxy integrated 3PCFs (red contours), the constraints remain unbiased, but they become visibly tighter. This shows that our conservative scale cuts are adequate to return unbiased results, while still letting our tree-level models explore the non-Gaussian information in the galaxy correlations to improve the parameter constraints. Relative to the galaxy 2PCFs constraints (blue), the addition of the galaxy integrated 3PCFs (red) improves the constraints on $\ln(10^{10}A_s)$, $b_{1}^{\text{\tiny{BIN1}}}$, $b_{1}^{\text{\tiny{BIN2}}}$ by approximately $30\%$, $20\%$, $25\%$, respectively. These improvements are associated to the breaking of degeneracies between $A_s$ and $b_1$ in the galaxy 2PCFs by the galaxy integrated 3PCFs. The galaxy integrated 3PCFs can also constrain the $b_2$ parameter, which is not possible with $\xi_{t}$, $\xi_{g}$ at tree-level.

The predictions of the tree-level models using the best-fitting parameters from this MCMC analysis are shown by the blue dashed curves in Fig.~\ref{fig:correlations}. As expected, they agree with the predictions for the fiducial parameters (red curves) on scales larger than our assumed scale cuts.

\subsection{Fisher forecasts for a DES-Y3-like survey}
\label{sec:Fisher}

We now investigate in the context of Fisher matrix forecasts the ability of combined 2PCFs and integrated 3PCFs to constrain cosmological parameters; from hereon we consider also the cosmic shear-only 2PCFs $\xi_{\pm}$ and integrated 3PCFs $\zeta_{a\pm}$. The Fisher information matrix $\mathbf{F}$ for a model vector $M$ depending on parameters $\pi$, assuming a constant data-covariance $\mathbf{C}$, is given by \cite{Tegmark1997}
\begin{equation} \label{eq:fisher_model_vector}
    F_{ij} = \left( \frac{\partial M(\boldsymbol{\pi})}{\partial \pi_{i}} \right)^{\mathrm{T}} \mathbf{C}^{-1} \left( \frac{\partial M(\boldsymbol{\pi})}{\partial \pi_{j}} \right),
\end{equation}
where $F_{ij}$ is an element of the matrix $\mathbf{F}$ associated with the parameters $\pi_{i}$ and $\pi_{j}$, and $\mathbf{C}^{-1}$ is the inverse data covariance matrix in Eq.~\eqref{eq:precision_hartlap}. The partial derivative of the model vector with respect to the parameter $\pi_{i}$ can be computed using a 2-point central difference:
\begin{equation}
    \frac{\partial M(\boldsymbol{\pi})}{\partial \pi_{i}} = \frac{M(\pi_{i} + \delta_{i}) - M(\pi_{i} - \delta_{i})}{2 \delta_{i}},
\end{equation}
where $\delta_{i}$ is a small change in the parameter $\pi_{i}$ around its fiducial value, and $M(\pi_{i} \pm \delta_{i})$ is the model vector computed at the changed parameter $\pi_{i} \pm \delta_{i}$ with all the other parameters fixed. We consider the following cosmological and baryonic feedback parameters $\boldsymbol{\pi}_{\text{cosmo}} = \{ \Omega_{\mathrm{cdm}}, \ln(10^{10}A_s), w_0, w_a, h, c_{\mathrm{\rm min}} \}$, where $w_0, w_a$ are the dynamical dark energy equation of state parameters (in the CPL parametrization $w(z) = w_0 + w_a z/(1+z)$ \cite{chevallier_2001}) and $c_{\mathrm{\rm min}}$ is a baryonic feedback parameter of the \verb|HMCODE| \cite{Mead2015} nonlinear matter power spectrum which enters in our modelling of the cosmic shear-only statistics $\xi_{\pm}$ and $\zeta_{a\pm}$. The fiducial values are $\boldsymbol{\pi}_0 = \{0.233, 3.089, -1, 0.0, 0.7, 3.13\}$; the cosmological parameters are the same as the T17 simulations, and the baryonic parameter is the gravity-only value as determined by Ref.~\cite{Mead2015}. When varying the $\Omega_{\mathrm{cdm}}$ parameter we keep the baryon density $\Omega_{\mathrm{b}}$ fixed, but adjust the dark energy density to keep the universe spatially flat. In addition, we also vary the galaxy bias $b_1, b_2$ and point-mass $\mathcal{M}$ parameters for both lens samples. The fiducial values of the galaxy bias terms are given in Tab.~\ref{tab:HOD_parameters}; the point-mass term fiducial values are assumed to be zero in our analysis. When varying $b_1$ we evaluate the tidal bias terms $b_{s^2}$ according to the co-evolution relation $b_{s^2}(b_1)$.

The parameter covariance matrix $\mathbf{C}_{\boldsymbol{\pi}}$ is given by the inverse of the Fisher matrix
\begin{equation} 
\label{eq:parameter_covariance_matrix}
    \mathbf{C}_{\boldsymbol{\pi}} = \mathbf{F}^{-1},
\end{equation}
which we use to forecast constraints on the model parameters. In our results below, we report the Fisher constraints on the cosmological parameters, baryonic feedback parameter and linear bias parameter $b_1$, marginalizing over the second-order bias $b_2$ and point-mass terms. We present results for three different combinations of 2- and 3-point statistics: (i) a 3$\times$2PCF-only analysis, labelled as $\{ \xi_{\pm}, \xi_{t}, \xi_{g} \}$ and shown in blue colour; (ii) the same, but adding the shear-only integrated 3PCF, labelled as $\{ \xi_{\pm}, \xi_{t}, \xi_{g}, \zeta_{a\pm} \}$ and shown in green; and (iii) using all of the statistics discussed in this paper combined, labelled as {\it all correlations} and shown in red. For each of these, we discuss three analysis setup cases:

\begin{figure}
\centering 
\includegraphics[width=\textwidth]{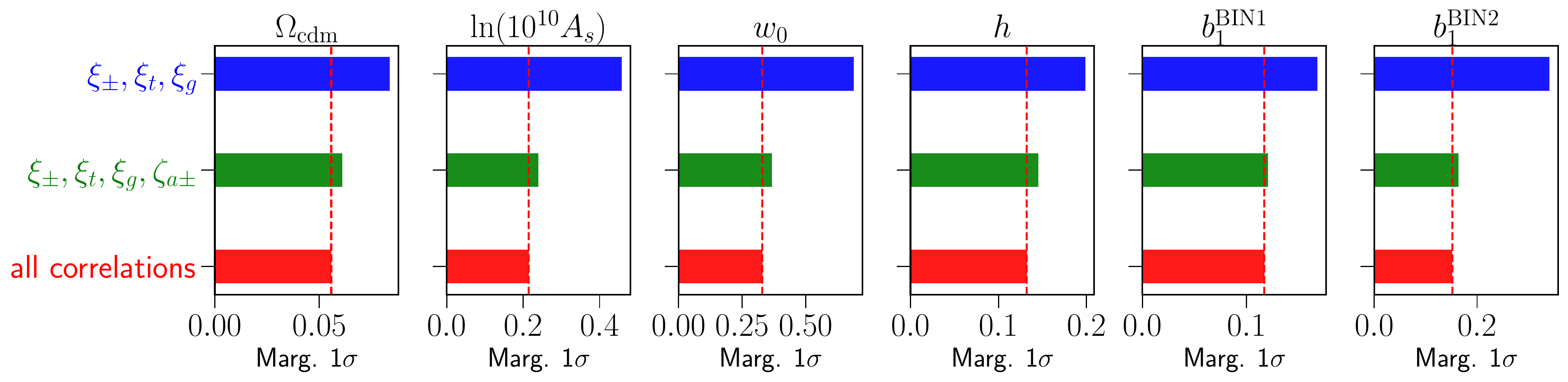}
\caption{\label{fig:Fisher_Omega_As_w0_h_b1_b1} Marginalized 1$\sigma$ Fisher constraints for analysis setup case A: constraints on the parameters $\{\Omega_{\rm cdm}, \ln(10^{10} A_s), w_0, h, b_{1}^{\text{\tiny{BIN1}}}, b_{1}^{\text{\tiny{BIN2}}} \}$ with scale cuts on all of the statistics. The columns are for the different parameters and the rows for different combinations of the 2PCFs and integrated 3PCFs. The second-order bias parameters $b_2$ and point-mass terms are marginalized over. The red dashed vertical lines serve as a guide to the eye for comparing the constraints from \textit{all correlations} with those from other combinations of $\xi$ and $\zeta$.}
\end{figure}

\begin{figure}
\centering 
\includegraphics[width=\textwidth]{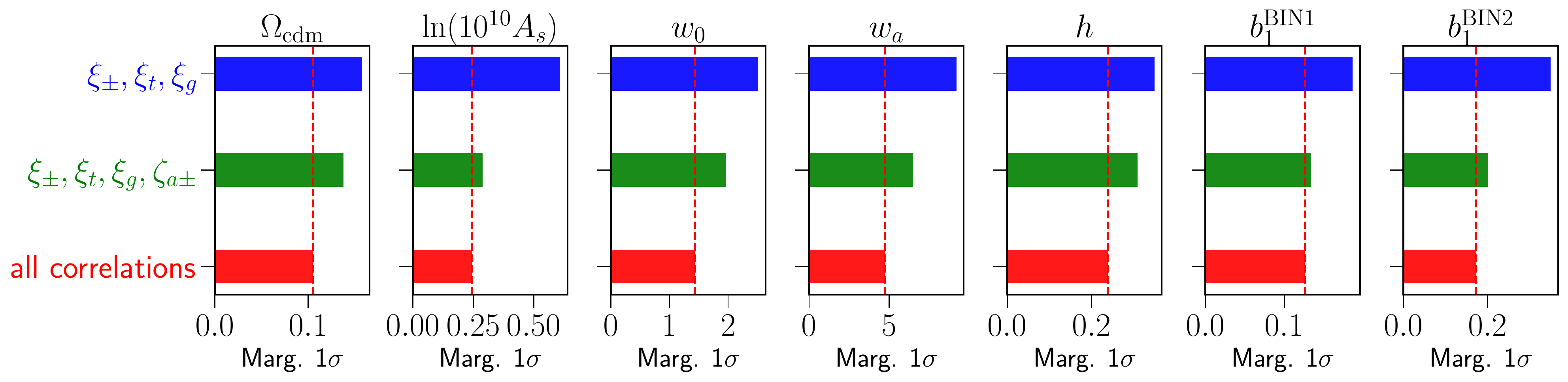}
\caption{\label{fig:Fisher_Omega_As_w0_wa_h_b1_b1} Same as Fig.~\ref{fig:Fisher_Omega_As_w0_h_b1_b1} but for analysis setup case B, in which $w_a$ is also a free parameter.}
\end{figure}

\begin{figure}
\centering 
\includegraphics[width=\textwidth]{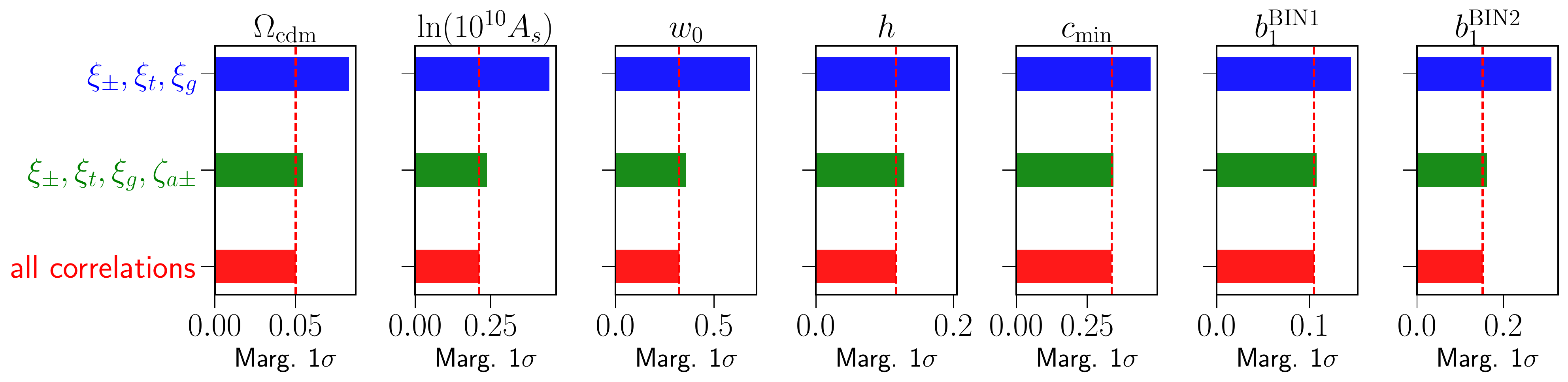}
\caption{\label{fig:Fisher_Omega_As_w0_h_cmin_b1_b1} Same as Fig.~\ref{fig:Fisher_Omega_As_w0_h_b1_b1} but for analysis setup case C: no scale cuts on the shear-only statistics $\xi_{\pm}, \zeta_{a\pm}$ (conservative scale cuts are still imposed on the galaxy correlations) and $c_{\rm min}$ is a free parameter.}
\end{figure}

\begin{itemize}

\item \underline{\it Case A}: constraints on the parameters $\{\Omega_{\mathrm{cdm}}, \ln(10^{10}A_s), w_0, h, b_{1}^{\text{\tiny{BIN1}}}, b_{1}^{\text{\tiny{BIN2}}}\}$, assuming scale cuts on all statistics (cf.~vertical lines in Fig.~\ref{fig:correlations}). Recall that the scale cuts on the cosmic shear-only statistics $\xi_{\pm}, \zeta_{a\pm}$ are the ones deemed as baryon-free, and so we keep $c_{\mathrm{\rm min}}$ fixed to the fiducial value (we also fix $w_a$). The 1$\sigma$ marginalized Fisher constraints are shown in Fig.~\ref{fig:Fisher_Omega_As_w0_h_b1_b1}. Relative to the 3$\times$2PCF constraints (blue), the addition of the shear-only integrated 3PCF $\zeta_{a\pm}$ (green) improves the constraints on $\{\Omega_{\mathrm{cdm}}, \ln(10^{10}A_s), w_0, h\}$ by $\{27, 48, 47, 27\}\%$, respectively. The addition of all other integrated 3PCFs involving the galaxy density field leads to further $\{9, 10, 10, 9\}\%$ improvements (red vs.~green). The constraints on the bias parameters $\{b_{1}^{\text{\tiny{BIN1}}}, b_{1}^{\text{\tiny{BIN2}}}\}$ are also improved by the integrated 3PCFs: $\{28, 52\}\%$ from the blue to the green and $\{3, 7\}\%$ from the green to the red constraints.

\item \underline{\it Case B} is the same as case A, but with $w_a$ treated as a free parameter; the constraints are shown in Fig.~\ref{fig:Fisher_Omega_As_w0_wa_h_b1_b1}. Relative to the 3$\times$2PCFs, $\zeta_{a\pm}$ leads to improvements of $\{13, 53, 22, 29, 11\}\%$ on $\{\Omega_{\mathrm{cdm}}, \ln(10^{10}A_s), w_0, w_a, h\}$ (green vs.~blue). The addition of all other galaxy integrated 3PCF correlations leads to further improvements of $\{24, 15, 27, 27, 23\}\%$. The same improvements for $\{b_{1}^{\text{\tiny{BIN1}}}, b_{1}^{\text{\tiny{BIN2}}}\}$ are $\{28, 42\}\%$ (green vs.~blue) and $\{6, 14\}\%$ (red vs.~green).

\item \underline{\it Case C} is the same as case A, but without any scale cuts on the shear-only statistics $\xi_{\pm}, \zeta_{a\pm}$ (we still apply the scale cuts on the galaxy correlations). Hence, in this case we also vary the baryonic feedback parameter $c_{\mathrm{\rm min}}$. We fix $w_a$ again to its fiducial value. The results are shown in Fig.~\ref{fig:Fisher_Omega_As_w0_h_cmin_b1_b1}. The $\zeta_{a\pm}$ statistic improves the constraints on $\{\Omega_{\mathrm{cdm}}, \ln(10^{10}A_s), w_0, h,  c_{\mathrm{\rm min}}\}$ by $\{34, 47, 47, 34, 27\}\%$, relative to 3$\times$2PCFs (green vs.~blue). The rest of the galaxy correlations can further improve these by $\{8, 10, 10, 9 ,2\}\%$ (red vs.~green). The same improvements for $\{b_{1}^{\text{\tiny{BIN1}}}, b_{1}^{\text{\tiny{BIN2}}}\}$ are $\{25, 48\}\%$ (green vs.~blue) and $\{3, 6\}\%$ (red vs.~green).

\end{itemize}

The main takeaway from Figs.~\ref{fig:Fisher_Omega_As_w0_h_b1_b1}, \ref{fig:Fisher_Omega_As_w0_wa_h_b1_b1}, \ref{fig:Fisher_Omega_As_w0_h_cmin_b1_b1} is that the bulk of the improvements from adding integrated 3PCF information to 3$\times$2PCF-only analyses comes from the addition of the cosmic shear-only integrated 3PCF $\zeta_{a\pm}$ ($20-40$\%; green bars), with the remainder of the integrated 3PCFs involving the galaxy density field adding smaller, but still significant improvements of approximately 10\% (red bars). The smaller improvements by these galaxy correlations must be interpreted however in light of the conservative range of scales we assumed for them; higher-order perturbation theory calculations of the galaxy-matter bispectrum valid on smaller scales may result in further information gains in cosmological parameter constraints even when marginalizing over larger number of bias parameters (see Ref.~\cite{Eggemeier_2021}). In addition, even for our conservative scale cuts, these galaxy correlations show the potential to already lead to improvements of up to $20\%$ in parameter constraints when $w_a$ is included as a free parameter. Besides cosmology, another advantage of the galaxy integrated 3PCFs is that they can tighten constraints not only on linear bias, but also higher-order galaxy bias parameters.

We note further that the improvements reported here are also tied to other analysis setup choices such as the parameter space, number of lens and source tomographic bins and galaxy number density. For example, the precise numbers may change when  increasing the number of tomographic bins in the analysis. In any case, we expect that the addition of the integrated 3PCFs will always help to lift parameter degeneracies present at the 2PCF level and lead generically to improved constraints.

\section{Summary and Conclusion}
\label{sec:conclusion}

The integrated 3-point function is a practical statistic that measures the correlation between the local 2-point correlation function and 1-point averages in patches of the survey, and which probes the squeezed-limit of the full 3-point function. In Refs.~\cite{Halder2021, Halder2022, Gong2023cosmology} this formalism has been developed for the case of the cosmic shear field, where the statistic is known as the integrated shear 3-point correlation function $\zeta_{a\pm}$. In this paper, we have extended the formalism to include the foreground galaxy distribution and its cross-correlations with the shear field, which results in 5 new integrated 3PCFs, $\{ \zeta_{at}, \zeta_{ag}, \zeta_{g\pm}, \zeta_{gt}, \zeta_{gg} \}$. In total, these 6 statistics describe the correlation between (i) three position-dependent 2PCFs, namely cosmic shear 2PCF $\xi_{\pm}$, tangential shear 2PCF $\xi_{t}$ and galaxy clustering 2PCF $\xi_{g}$, and (ii) two average 1-point statistics, namely the lensing aperture mass $M_a$ and the average foreground galaxy density $M_g$ (see Fig.~\ref{fig:6xi3PCFs_illustrations} for an illustration). This forms a set of higher-order galaxy and shear statistics that can be readily measured from survey data, and thus be used to improve cosmological parameter constraints relative to standard analyses based on 2PCFs alone.

The main objectives of our work were to:

\begin{enumerate}
\item Introduce the integrated 3PCFs involving the galaxy and cosmic shear fields, and put forward theory model predictions to evaluate them (Sec.~\ref{sec:zeta} and App.~\ref{app:3D_spectra}). For the shear-only $\zeta_{a\pm}$ statistic we use the response approach to perturbation theory which is accurate in the nonlinear regime. For all other statistics involving the galaxy density field we use tree-level perturbation theory.

\item Identify the regime of validity of the theory predictions against measurements from realistic mock galaxy and lensing simulated data (see Secs.~\ref{sec:simulations}, \ref{sec:measurements} and \ref{sec:validation_MCMC}). We considered a DES Y3-like setup with two foreground lens galaxy bins and a single source lensing bin.

\item Investigate the improvement in cosmological, baryonic and galaxy bias parameter constraints from adding the integrated 3PCFs to standard analyses based on global 3$\times$2PCFs.

\end{enumerate}
Our main results can be summarized as follows: 

\begin{itemize}

\item Concerning the integrated 3PCFs involving the galaxy density field that are new to this work, we find that even when restricting to large angular scales (with conservative scale cuts) where our tree-level perturbation theory models are valid (cf.~Fig.~\ref{fig:correlations}), these higher-order statistics can still lead to improvements in parameter constraints. In an MCMC constraint analysis on the cosmological parameter $A_s$ and the galaxy bias parameters $b_1$ and $b_2$, we found that the corresponding fiducial parameter values can be recovered within $68\%$ credible intervals and the addition of the integrated 3PCFs $\zeta_{at}, \zeta_{ag}, \zeta_{g\pm}, \zeta_{gt}, \zeta_{gg}$, could lead to $20-30\%$ improvements over the constraints from the galaxy 2PCFs $\xi_{t}, \xi_{g}$.

\item Using Fisher matrix forecasts for a DES-Y3-like survey, we find that the addition of the shear-only integrated 3PCF $\zeta_{a\pm}$ can lead to $20-40\%$ improvements on the constraints of parameters like $\Omega_{\rm cdm}, \ln(10^{10}A_s), w_0, h, b_1$, relative to the standard analysis with the 3$\times$2PCFs alone (cf.~green vs.~blue in Figs.~\ref{fig:Fisher_Omega_As_w0_h_b1_b1} and \ref{fig:Fisher_Omega_As_w0_h_cmin_b1_b1}).

\item The addition of the remainder integrated 3PCFs involving the galaxy density field, even when restricted to conservatively large scales, can further improve the constraints by $\sim 10\%$ (cf.~red vs.~green in Figs.~\ref{fig:Fisher_Omega_As_w0_h_b1_b1} and \ref{fig:Fisher_Omega_As_w0_h_cmin_b1_b1}). These improvements depend however on the specific analysis setup: for example, in constraints where the dynamical dark energy parameter $w_a$ is free, these improvements can become $\sim 15 - 25\%$ (cf. Fig.~\ref{fig:Fisher_Omega_As_w0_wa_h_b1_b1}).

\end{itemize}

These results are encouraging and motivate further developments on the theory modelling front. This includes more accurate modelling of the galaxy-matter bispectrum on smaller scales (e.g.~one-loop bispectrum \cite{Eggemeier_2021}) to utilize the higher signal-to-noise $\zeta$ data points currently excluded in our analysis due to conservative scale cuts, modelling of redshift space distortions \cite{Leicht_2021} and wide-angle effects \cite{Pardede_2023}, as well as observational systematic effects such as galaxy intrinsic alignments \cite{Schmitz_2018}, photometric redshift uncertainty and shear calibration bias. In particular, it would be interesting to generalize the work of Ref.~\cite{Gong2023cosmology} on the integrated shear 3PCF, who investigated the optimal size of apertures for measuring $\zeta_{a\pm}$ and developed an end-to-end $\zeta_{a\pm}$ MCMC analysis pipeline, to the case of the integrated 3PCFs involving the galaxy density field. Jointly analysing the 3$\times$2PCFs and the integrated 3PCFs in cases beyond the standard cosmological parameters such as in searches for primordial non-Gaussianity and massive neutrinos using galaxy imaging and CMB lensing surveys would also be interesting avenues to explore in future works.

Overall, our results indicate that there is important cosmological information in integrated 3-point correlation functions involving the galaxy field and its cross-correlations with cosmic shear. Crucially, these statistics can be straightforwardly measured using existing and well-tested estimators for 1- and 2-point statistics, enabling the exploration of 3-point function information in current galaxy imaging surveys such as DES, and in the future using Euclid and Vera Rubin LSST data.

\acknowledgments

AH would like to thank Jonathan Blazek, Eiichiro Komatsu, Elisabeth Krause, Marilena LoVerde and Jochen Weller for helpful discussions. We acknowledge support from the Excellence Cluster ORIGINS which is funded by the Deutsche Forschungsgemeinschaft (DFG, German Research Foundation) under Germany's Excellence Strategy - EXC-2094-390783311. Some of the numerical calculations have been carried out on the ORIGINS computing facilities of the Computational Center for Particle and Astrophysics (C2PAP). The results in this paper have been derived using the following publicly available libraries and software packages: \verb|healpy| \cite{Zonca2019}, \verb|TreeCorr| \cite{Jarvis_2004}, \verb|emcee| \cite{Foreman_Mackey_2013}, \verb|CLASS| \cite{lesgourgues2011}, \verb|NumPy| \cite{harris2020numpy}, \verb|matplotlib| \cite{Hunter:2007}, and \verb|ChainConsumer| \cite{Hinton2016}.

\paragraph{Data availability}

The numerical data underlying the analysis of this paper may be shared upon reasonable request to the authors. 

\appendix

\section{Projected galaxy and weak lensing fields}

\label{app:2D_projected_fields}

In this appendix we outline the background behind the cosmic shear and the projected galaxy density contrast fields we consider in this work.

\subsection{Tangential and cross components of the shear field}

The complex weak lensing shear $\gamma(\vtheta) = \gamma_1(\vtheta) + i \gamma_2(\vtheta)$ has components $\gamma_1$ and $\gamma_2$ specified in a given Cartesian frame. However, one is free to rotate the coordinates by any arbitrary angle $\beta$. With respect to this reference rotation angle $\beta$, one defines the \textit{rotated shear}
\begin{equation}
\gamma_{\beta}(\vtheta) \equiv -e^{-2i\beta} \gamma(\vtheta)  = -e^{-2i\beta}\big[\gamma_1(\vtheta) + i \gamma_2(\vtheta)\big],
\end{equation}
and correspondingly the \textit{tangential} ($t$) and \textit{cross} ($\times$) components of the shear at position $\vtheta$ w.r.t.~the reference rotation angle $\beta$ as
\begin{equation}
    \gamma_{\beta}(\vtheta) \equiv \gamma_{t}(\vtheta,\beta) + i \gamma_{\times}(\vtheta,\beta) = -e^{-2i\beta}\big[\gamma_1(\vtheta) + i \gamma_2(\vtheta)\big].
\end{equation}
In particular, given a pair of points $\vtheta$ and $\v\vartheta$ on the sky separated by $\valpha \equiv \v\vartheta - \vtheta$, one can write the tangential and cross components of the shear along $\beta = \phi_{\valpha}$ (where $\phi_{\valpha}$ is polar angle of $\valpha$).

\subsection{Projected galaxy number density contrast field}

The projected number of galaxies $N(\vtheta)$ at position $\vtheta$ can be written as a line-of-sight projection of the three-dimensional galaxy number density $n_g^{\mathrm{3D}}(\vx, \tau)$ along the comoving radial coordinate $\chi$:
\begin{equation}
\begin{split}
    N(\vtheta) & = \int \mathrm{d} \chi \; \frac{\mathrm{d} V}{\mathrm{d} \chi} \; n_g^{\mathrm{3D}}(\vx, \tau)
    \equiv \Bar{N} \; [ 1 + \delta_{g}^{\mathrm{2D}}(\vtheta) ],
\end{split}
\end{equation}
where $\vx = (\chi\vtheta, \chi)$, $\frac{\mathrm{d} V}{\mathrm{d} \chi}(\chi)$ is the cosmological volume element (which is survey specific and for example for the whole spherical sky is $\frac{\mathrm{d} V}{\mathrm{d} \chi} (\chi)= 4\pi \chi^2$) and $ \Bar{N} \equiv \int \mathrm{d} \chi \; \frac{\mathrm{d} V}{\mathrm{d} \chi} \; \Bar{n}_g^{\mathrm{3D}}\left(\tau(\chi)\right)$ is the average number count of galaxies.  The projected galaxy number density contrast field $\delta_{g}^{\mathrm{2D}}(\vtheta)$ is defined as
\begin{equation}
\begin{split}
    \delta_{g}^{\mathrm{2D}}(\vtheta) \equiv \frac{1}{\Bar{N}} \int \mathrm{d} \chi \; \frac{\mathrm{d} V}{\mathrm{d} \chi} \; \Bar{n}_g^{\mathrm{3D}}(\tau) \; \delta_{g}^{\mathrm{3D}}(\vx,\tau) = \int \mathrm{d} \chi \; q_g(\chi) \; \delta_{g}^{\mathrm{3D}}(\vx,\tau),
\end{split}
\end{equation}
where in the second equality we have identified the galaxy projection kernel as $q_g(\chi) = \frac{1}{\Bar{N}} \frac{\mathrm{d} V}{\mathrm{d} \chi}  \Bar{n}_g^{\mathrm{3D}}(\tau(\chi))$. 

\section{HOD expressions for galaxy bias}

\label{app:HOD}

The halo model and halo occupation distribution (HOD) approach offer a useful framework to make predictions for the galaxy bias parameters defined in Eq.~\eqref{eq:bias_expansion_upto_second_order_RS}. The halo model assumes that galaxies reside inside dark matter halos with some mass. In this HOD approach the effective global number density of galaxies hosted within halos of mass $M_h \in [M_{h,min},M_{h,max}]$ is given by
\begin{equation}
    \bar{n}_g^{\mathrm{3D}}(\tau) = \int_{M_{h,min}}^{M_{h,max}} \mathrm{d}M_h \; \frac{\mathrm{d}\bar{n}_h^{\mathrm{3D}}}{\mathrm{d}M_h}(M_h,\tau) \; \bar{N}_g(M_h,\tau),
\end{equation}
where $\mathrm{d}\bar{n}_h^{\mathrm{3D}} / \mathrm{d}M_h$ is the global halo mass function (number density of dark matter halos in an infinitesimal mass bin $\mathrm{d}M_h$ around halos of mass $M_h$) and $\bar{N}_g(M_h,\tau)$ is the expected number of galaxies  residing inside dark matter halos of mass $M_h$ at time $\tau$. The galaxy bias parameters $b_{\mathcal{O}}$ are in turn expressed as \cite{Voivodic_2021}
\begin{equation}
\label{eq:galaxy_bias_HOD}
    b_{\mathcal{O}}(\tau) = \frac{1}{\bar{n}_g^{\mathrm{3D}}(\tau)}\int_{M_{h,min}}^{M_{h,max}} \mathrm{d}M_h \; \frac{\mathrm{d}\bar{n}_h^{\mathrm{3D}}}{\mathrm{d}M_h}(M_h,\tau) \; \bar{N}_g(M_h,\tau) \;  \big[ b_{\mathcal{O},h}(M_h,\tau) + R_{\mathcal{O},{\bar{N}_g}}(M_h,\tau) \big],
\end{equation}
where $b_{\mathcal{O},h}$ is the bias parameter of dark matter halos of mass $M_h$, and $R_{\mathcal{O},{\bar{N}_g}}(M_h,\tau)$ is called the {\it response function} of $\bar{N}_g$ for large-scale perturbations $\mathcal{O}$. This response function describes the modulation of $\bar{N}_g$ by the $\mathcal{O}$ perturbations, in the same way that the bias parameters $b_{\mathcal{O}, h}$ describe the modulation of the halo mass function. In our HOD catalogues, we have assumed $\bar{N}_g$ to be the same everywhere inside the simulation box irrespective of the local density and tidal field values, which corresponds to assuming $R_{\mathcal{O},{\bar{N}_g}} = 0$. In order to get an effective bias $b_{\mathcal{O}}^i$ parameter of a galaxy sample in a tomographic bin $i$ we take the expectation value of the galaxy bias parameter $b_{\mathcal{O}}(z)$ over the redshift distribution $p^i(z)$ of the bin:
\begin{equation}
b_{\mathcal{O}}^i = \int \mathrm{d}z \; p^i(z) b_{\mathcal{O}}(z).
\end{equation}

\section{Power spectra and bispectra of galaxy and matter density fields}

\label{app:3D_spectra}

In this appendix we discuss the leading-order (tree-level) standard perturbation theory (PT) models of the 3D galaxy-matter power spectra and bispectra used in our work. We can write the Fourier transform of  Eq.~\eqref{eq:bias_expansion_upto_second_order_RS} as
\begin{equation}
\begin{split}
    \delta^{\mathrm{3D}}_{g}(\boldsymbol{k}, \tau) & = b_1(\tau) \delta(\boldsymbol{k}, \tau) + \frac{1}{2} \int \frac{\mathrm{d}^3 \boldsymbol{q}}{(2\pi)^3} \; \delta(\boldsymbol{q}, \tau) \delta(\boldsymbol{k} - \boldsymbol{q}, \tau) \Big( b_2(\tau) + b_{s^2}(\tau)S_2(\boldsymbol{q}, \boldsymbol{k} - \boldsymbol{q}) \Big) \\
    & \qquad + \left[ \epsilon(\boldsymbol{k}, \tau) + \int \frac{\mathrm{d}^3 \boldsymbol{q}}{(2\pi)^3} \; \epsilon_{\delta}(\boldsymbol{q}, \tau) \delta(\boldsymbol{k} - \boldsymbol{q}, \tau) \right],
\end{split}
\end{equation}
where $\epsilon_{\mathcal{O}}$ are random Poisson variables with vanishing expectation values that are uncorrelated with the density fields. The term $S_2(\boldsymbol{k}, \boldsymbol{q})$ is the operator which generates the Fourier representation of the square of the tidal tensor $K^2$ \cite{Baldauf_2012}
\begin{equation}
    S_2(\boldsymbol{k}, \boldsymbol{q}) = \frac{(\boldsymbol{k} \cdot \boldsymbol{q})^2}{(kq)^2} - \frac{1}{3} \; .
\end{equation}
In the equations that follow we drop the time $\tau$ from the arguments to ease the notation.

\subsection{3D power spectra}

The 3D matter power spectrum $P^{\mathrm{3D}}_{mm}(k)$, the galaxy-matter cross-power spectrum $P^{\mathrm{3D}}_{gm}(k)$, and the galaxy power spectrum $P^{\mathrm{3D}}_{gg}(k)$ are defined as
\begin{subequations}
\begin{align}
\langle \delta^{\mathrm{3D}}_{m}(\boldsymbol{k}) \delta^{\mathrm{3D}}_{m}(\boldsymbol{k}') \rangle &= (2\pi)^3 \delta_D(\boldsymbol{k}+\boldsymbol{k}')P^{\mathrm{3D}}_{mm}(k), \\
\langle \delta^{\mathrm{3D}}_{g}(\boldsymbol{k}) \delta^{\mathrm{3D}}_{m}(\boldsymbol{k}') \rangle &= (2\pi)^3 \delta_D(\boldsymbol{k}+\boldsymbol{k}')P^{\mathrm{3D}}_{gm}(k), \\
\langle \delta^{\mathrm{3D}}_{g}(\boldsymbol{k}) \delta^{\mathrm{3D}}_{g}(\boldsymbol{k}') \rangle &= (2\pi)^3 \delta_D(\boldsymbol{k}+\boldsymbol{k}')P^{\mathrm{3D}}_{gg}(k).
\end{align}
\end{subequations}
At tree-level perturbation theory these are given by
\begin{subequations}
\begin{align}
P^{\mathrm{3D}}_{mm}(k) &= P^{\mathrm{3D}}_{\mathrm{lin}}(k), \\
P^{\mathrm{3D}}_{gm}(k) &= b_1 P^{\mathrm{3D}}_{\mathrm{lin}}(k), \\
P^{\mathrm{3D}}_{gg}(k) &= b_1^2 P^{\mathrm{3D}}_{\mathrm{lin}}(k) + P^{\mathrm{3D}}_{\epsilon\epsilon}(k),
\end{align}
\end{subequations}
where $P^{\mathrm{3D}}_{\epsilon\epsilon}$ is the power spectrum of the stochastic field $\eps$ and $P^{\mathrm{3D}}_{\mathrm{lin}}$ is the linear matter power spectrum which scales with the amplitude of the primordial scalar perturbations $A_s$, i.e.~$P^{\mathrm{3D}}_{\mathrm{lin}} \propto A_s$.

\subsection{3D bispectra}

The 3D matter bispectrum $B^{\mathrm{3D}}_{mmm}$, the galaxy-matter-matter bispectrum $B^{\mathrm{3D}}_{gmm}$, the galaxy-galaxy-matter bispectrum $B^{\mathrm{3D}}_{ggm}$, and the galaxy bispectrum $B^{\mathrm{3D}}_{ggg}$ are defined as
\begin{subequations}
\begin{align}
 \langle \delta^{\mathrm{3D}}_{m}(\boldsymbol{k}_1) \delta^{\mathrm{3D}}_{m}(\boldsymbol{k}_2) \delta^{\mathrm{3D}}_{m}(\boldsymbol{k}_3) \rangle &= (2\pi)^3 \delta_D(\boldsymbol{k}_1+\boldsymbol{k}_2+\boldsymbol{k}_3)B^{\mathrm{3D}}_{mmm}(\boldsymbol{k}_1,\boldsymbol{k}_2,\boldsymbol{k}_3), \\
 \langle \delta^{\mathrm{3D}}_{g}(\boldsymbol{k}_1) \delta^{\mathrm{3D}}_{m}(\boldsymbol{k}_2) \delta^{\mathrm{3D}}_{m}(\boldsymbol{k}_3) \rangle &= (2\pi)^3 \delta_D(\boldsymbol{k}_1+\boldsymbol{k}_2+\boldsymbol{k}_3)B^{\mathrm{3D}}_{gmm}(\boldsymbol{k}_1,\boldsymbol{k}_2,\boldsymbol{k}_3), \\
 \langle \delta^{\mathrm{3D}}_{g}(\boldsymbol{k}_1) \delta^{\mathrm{3D}}_{g}(\boldsymbol{k}_2) \delta^{\mathrm{3D}}_{m}(\boldsymbol{k}_3) \rangle &= (2\pi)^3 \delta_D(\boldsymbol{k}_1+\boldsymbol{k}_2+\boldsymbol{k}_3)B^{\mathrm{3D}}_{ggm}(\boldsymbol{k}_1,\boldsymbol{k}_2,\boldsymbol{k}_3), \\ 
 \langle \delta^{\mathrm{3D}}_{g}(\boldsymbol{k}_1) \delta^{\mathrm{3D}}_{g}(\boldsymbol{k}_2) \delta^{\mathrm{3D}}_{g}(\boldsymbol{k}_3) \rangle &= (2\pi)^3 \delta_D(\boldsymbol{k}_1+\boldsymbol{k}_2+\boldsymbol{k}_3)B^{\mathrm{3D}}_{ggg}(\boldsymbol{k}_1,\boldsymbol{k}_2,\boldsymbol{k}_3).
\end{align}
\end{subequations}
At tree-level perturbation theory these are expressed as \cite{biasreview,Leicht_2021}:
\begin{subequations}
\begin{align}
B^{\mathrm{3D}}_{mmm}(\boldsymbol{k}_1,\boldsymbol{k}_2,\boldsymbol{k}_3) &= 2 F_2(\boldsymbol{k}_1,\boldsymbol{k}_2)P^{\mathrm{3D}}_{\mathrm{lin}}(k_1)P^{\mathrm{3D}}_{\mathrm{lin}}(k_2) + 2 F_2(\boldsymbol{k}_3,\boldsymbol{k}_1)P^{\mathrm{3D}}_{\mathrm{lin}}(k_3,)P^{\mathrm{3D}}_{\mathrm{lin}}(k_1) \nonumber \\
    & \qquad + 2 F_2(\boldsymbol{k}_2,\boldsymbol{k}_3)P^{\mathrm{3D}}_{\mathrm{lin}}(k_2)P^{\mathrm{3D}}_{\mathrm{lin}}(k_3) \equiv B^{\mathrm{3D}}_{tree}(\boldsymbol{k}_1,\boldsymbol{k}_2,\boldsymbol{k}_3), \\
    B^{\mathrm{3D}}_{gmm}(\boldsymbol{k}_1,\boldsymbol{k}_2,\boldsymbol{k}_3) &= b_1B^{\mathrm{3D}}_{tree}(\boldsymbol{k}_1,\boldsymbol{k}_2,\boldsymbol{k}_3) + \Big( b_2 + b_{s^2}S_2(\boldsymbol{k}_2,\boldsymbol{k}_3) \Big) P^{\mathrm{3D}}_{\mathrm{lin}}(k_2)P^{\mathrm{3D}}_{\mathrm{lin}}(k_3), \label{eq:B_3D_gmm} \\
B^{\mathrm{3D}}_{ggm}(\boldsymbol{k}_1,\boldsymbol{k}_2,\boldsymbol{k}_3) &= b_1^2 B^{\mathrm{3D}}_{tree}(\boldsymbol{k}_1,\boldsymbol{k}_2,\boldsymbol{k}_3) + b_1 \Big( b_2 + b_{s^2}S_2(\boldsymbol{k}_3,\boldsymbol{k}_1) \Big) P^{\mathrm{3D}}_{\mathrm{lin}}(k_3)P^{\mathrm{3D}}_{\mathrm{lin}}(k_1) \nonumber \\ 
    & \qquad + b_1\Big( b_2 + b_{s^2}S_2(\boldsymbol{k}_2,\boldsymbol{k}_3) \Big) P^{\mathrm{3D}}_{\mathrm{lin}}(k_2)P^{\mathrm{3D}}_{\mathrm{lin}}(k_3) + 2 P^{\mathrm{3D}}_{\epsilon\epsilon_{\delta}}(k_3)P^{\mathrm{3D}}_{\mathrm{lin}}(k_3), \label{eq:B_3D_ggm} \\
B^{\mathrm{3D}}_{ggg}(\boldsymbol{k}_1,\boldsymbol{k}_2,\boldsymbol{k}_3) & = b_1^3 B^{\mathrm{3D}}_{tree}(\boldsymbol{k}_1,\boldsymbol{k}_2,\boldsymbol{k}_3) + b_1^2 \Big( b_2 + b_{s^2}S_2(\boldsymbol{k}_1,\boldsymbol{k}_2) \Big) P^{\mathrm{3D}}_{\mathrm{lin}}(k_1)P^{\mathrm{3D}}_{\mathrm{lin}}(k_2) \nonumber  \\ 
    & \qquad + b_1^2 \Big( b_2 + b_{s^2}S_2(\boldsymbol{k}_3,\boldsymbol{k}_1) \Big) P^{\mathrm{3D}}_{\mathrm{lin}}(k_3)P^{\mathrm{3D}}_{\mathrm{lin}}(k_1) \nonumber  \\ 
    & \qquad + b_1^2 \Big( b_2 + b_{s^2}S_2(\boldsymbol{k}_2,\boldsymbol{k}_3) \Big) P^{\mathrm{3D}}_{\mathrm{lin}}(k_2)P^{\mathrm{3D}}_{\mathrm{lin}}(k_3) \nonumber  \\ 
    & \qquad + 2 b_1 \Big( P^{\mathrm{3D}}_{\epsilon\epsilon_{\delta}}(k_1)P^{\mathrm{3D}}_{\mathrm{lin}}(k_1) + P^{\mathrm{3D}}_{\epsilon\epsilon_{\delta}}(k_2)P^{\mathrm{3D}}_{\mathrm{lin}}(k_2) + P^{\mathrm{3D}}_{\epsilon\epsilon_{\delta}}(k_3)P^{\mathrm{3D}}_{\mathrm{lin}}(k_3) \Big) \nonumber  \\
    & \qquad + B^{\mathrm{3D}}_{\epsilon\epsilon\epsilon}(k_1, k_2, k_3), \label{eq:B_3D_ggg}
\end{align}
\end{subequations}
where $F_2$ is the second-order gravitational mode-coupling kernel. The tree-level matter bispectrum $B^{\mathrm{3D}}_{tree}$ scales differently with $A_s$ compared to the linear matter power spectrum, i.e.~$B^{\mathrm{3D}}_{tree} \propto (P^{\mathrm{3D}}_{\mathrm{lin}})^2 \propto A_s^2$. Further, the galaxy-matter bispectra scale differently with galaxy bias compared to the galaxy-matter power spectra. The $P^{\mathrm{3D}}_{\epsilon\epsilon}$, $P^{\mathrm{3D}}_{\epsilon\epsilon_{\delta}}$ and $B^{\mathrm{3D}}_{\epsilon\epsilon\epsilon}$ are the power- and bi-spectra of the stochastic fields.  Under the assumption of Poisson statistics for them, it follows that \cite{biasreview, Leicht_2021}:
\begin{equation}
P^{\mathrm{3D}}_{\epsilon\epsilon} = \frac{1}{\Bar{n}_g^{\rm 3D}}\ , \qquad P^{\mathrm{3D}}_{\epsilon\epsilon_{\delta}} = \frac{b_1}{2\Bar{n}_g^{\rm 3D}}\ , \qquad B^{\mathrm{3D}}_{\epsilon\epsilon\epsilon} = \frac{1}{(\Bar{n}_g^{\rm 3D})^2}  \ .
\end{equation}
We note that to evaluate $B^{\mathrm{3D}}_{gmm}(\boldsymbol{k}_1, \boldsymbol{k}_2, \boldsymbol{k}_3)$ or $B^{\mathrm{3D}}_{mgm}(\boldsymbol{k}_1, \boldsymbol{k}_2, \boldsymbol{k}_3)$ it is important to associate the correct ordering of wavevectors $\boldsymbol{k}_i$ to the respective galaxy density \textit{`g'} parts of the correlations. For example in $B^{\mathrm{3D}}_{gmm}(\boldsymbol{k}_1, \boldsymbol{k}_2, \boldsymbol{k}_3)$, the $\boldsymbol{k}_1$ mode is associated to $\delta_g$, whereas in $B^{\mathrm{3D}}_{mgm}(\boldsymbol{k}_1, \boldsymbol{k}_2, \boldsymbol{k}_3)$ it is instead the mode $\boldsymbol{k}_2$. One thus needs to alter the wavevector arguments in Eq.~\eqref{eq:B_3D_gmm} accordingly when calculating $B^{\mathrm{3D}}_{mgm}$. Similar considerations hold when there are two instances of $\delta_g$, e.g.~in $B^{\mathrm{3D}}_{ggm}(\boldsymbol{k}_1, \boldsymbol{k}_2, \boldsymbol{k}_3)$ and $B^{\mathrm{3D}}_{mgg}(\boldsymbol{k}_1, \boldsymbol{k}_2, \boldsymbol{k}_3)$.

\section{Point-mass terms in tangential shear 2PCFs and integrated 3PCFs}
\label{app:point_mass}

The \textit{mean} 3D matter density at position $\vx + \vr$ in the presence of a galaxy at $\vx$ can be written as \cite{Peebles1980}
\begin{equation}
    \rho_{m}^{\rm 3D}\left(\vx + \vr \ | \ n_{g}^{\rm 3D}(\vx) \right) \equiv \bar{\rho}_{m}^{\rm 3D}[1+\xi_{gm}(r)],
\end{equation}
where $\xi_{gm}(r) \equiv \langle \delta_{g}^{\rm 3D}(\vx) \delta_{m}^{\rm 3D}(\vx + \vr) \rangle$ is the 3D galaxy-matter 2-point correlation function (we have assumed statistical homogeneity and isotropy of the Universe).  The projected surface mass density on a 2D plane at position $\v R$ around a galaxy located at the origin is given by
\begin{equation}
    \Sigma_{gm}^{\rm 2D}(\v R) \equiv \int_0^{\chi_{\rm lim}} {\rm d} \chi \ \rho_{m}^{\rm 3D}\left(\vr \ | \ n_{g}^{\rm 3D}(\vx = \v0) \right),
\end{equation}
where $\vr = \left[\v R = (R, \phi), \chi \right]$ is expressed in cylindrical coordinates. The angle-averaged projected surface mass density at a distance $R$ from the galaxy density then reads
\begin{equation}
    \Sigma_{gm}(R) = \int_0^{\chi_{\rm lim}} {\rm d} \chi \ \bar{\rho}_{m}^{\rm 3D} \left[ 1+\xi_{gm}(\sqrt{R^2 + \chi^2}) \right],
\end{equation}
where $r = \sqrt{R^2 + \chi^2}$. The average surface mass density within the disc of radius $R$ around the galaxy position is given by
\begin{equation}
    \Bar{\Sigma}_{gm}(R) = \frac{2}{R^2}\int_0^{R} {\rm d} R' \ R' \ \Sigma_{gm}(R').
\end{equation}
The tangential shear signal is effectively a measure of the \textit{excess surface mass density} $\Delta \Sigma_{gm}$, which is defined as
\begin{equation}
\label{eq:excess_surface_mass}
\begin{split}
    \Delta \Sigma_{gm}(R) & = \Bar{\Sigma}_{gm}(R) - \Sigma_{gm}(R) \\
    & = \frac{2\bar{\rho}_{m}^{\rm 3D}}{R^2} \int_0^{\chi_{\rm lim}} {\rm d} \chi \ \int_0^{R} {\rm d} R' \ R' \ \xi_{gm}(\sqrt{R'^2 + \chi^2}) - \bar{\rho}_{m}^{\rm 3D} \int_0^{\chi_{\rm lim}} {\rm d} \chi \ \xi_{gm}(\sqrt{R^2 + \chi^2}).
\end{split}
\end{equation}
This equation shows that the result at a given $R$ gets contributions from $\xi_{gm}$ on all distance scales below it through the integral $\int_0^{R} {\rm d} R'$, including scales where perturbation theory breaks down. To circumvent the problem and be able to make predictions for the tangential shear signal we write the galaxy-matter correlation function as
\begin{equation}
\xi_{gm}(r) = \xi_{gm}^{\rm PT}(r) + \xi_{gm}^{\rm res}(r),
\end{equation}
where $\xi_{gm}^{\rm res}(r)$ is a residual term that is only nonzero for $r < r_{\rm min}$, with $r_{\rm min}$ denoting the scale below which perturbation theory breaks down. This way Eq.~(\ref{eq:excess_surface_mass}) can be written as
\begin{equation}
\begin{split}
    \Delta \Sigma_{gm}(R > R_{\rm min}) & = \Delta \Sigma_{gm}^{\rm PT}(R > R_{\rm min}) + \Delta \Sigma_{gm}^{\rm res}(R > R_{\rm min}), 
\end{split}
\end{equation}
where
\begin{equation}\label{eq:DsigmaPT}
    \Delta \Sigma_{gm}^{\rm PT}(R > R_{\rm min}) = \frac{2\bar{\rho}_{m}^{\rm 3D}}{R^2} \int_0^{\chi_{\rm lim}} {\rm d} \chi \ \int_0^{R} {\rm d} R' \ R' \ \xi_{gm}^{\rm PT}(\sqrt{R'^2 + \chi^2}) - \bar{\rho}_{m}^{\rm 3D} \int_0^{\chi_{\rm lim}} {\rm d} \chi \ \xi_{gm}^{\rm PT}(\sqrt{R^2 + \chi^2}),
\end{equation}
and 
\begin{equation}\label{eq:Dsigmares}
\begin{split}
    \Delta \Sigma_{gm}^{\rm res}(R > R_{\rm min}) & = \frac{2\bar{\rho}_{m}^{\rm 3D}}{R^2} \int_0^{\chi_{\rm lim}} {\rm d} \chi \ \int_0^{R_{\rm min}} {\rm d} R' \ R' \ \xi_{gm}^{\rm res}(\sqrt{R'^2 + \chi^2}),
\end{split}
\end{equation}
where we have used $\xi_{gm}^{\rm res}(R > R_{\rm min}) = 0$. Note that although Eq.~(\ref{eq:DsigmaPT}) integrates a perturbation theory model down to scales where it is not valid, the corrections to this when studying scales $R > R_{\rm min}$ are automatically absorbed by the residual contribution of Eq.~\eqref{eq:Dsigmares}; in the thin-lens approximation (where the lens mass is sharply concentrated around $\chi_l$, the comoving distance of the lens galaxies), the upper integration limit is given by $R_{\rm min} = \sqrt{r_{\rm min}^2 - \chi_l^2}$.

The observed tangential shear signal is related to the surface mass density via:
\begin{equation}
    \xi_t(\alpha) \equiv \frac{\Delta \Sigma_{gm}(R = \chi_l \alpha)}{\Sigma_{crit}},
\end{equation}
where 
\begin{equation}
    \Sigma_{crit} = \frac{4 \pi G}{c^2} \frac{(\chi_s - \chi_l) \chi_l}{\chi_s},
\end{equation}
and $\chi_s$ is the comoving distance of the source galaxies. This signal can thus be decomposed as 
\begin{equation}
\begin{split}
    \xi_{t}(\alpha > \alpha_{\rm min}) & = \xi_{t}^{\rm PT}(\alpha > \alpha_{\rm min}) + \xi_{t}^{\rm res}(\alpha > \alpha_{\rm min}),
\end{split}
\end{equation}
where
\begin{equation}
    \xi_{t}^{\rm PT}(\alpha > \alpha_{\rm min}) = \frac{\Delta \Sigma_{gm}^{\rm PT}(R = \chi_l \alpha)}{\Sigma_{crit}},
\end{equation}
and
\begin{equation}
\begin{split}
    \xi_{t}^{\rm res}(\alpha > \alpha_{\rm min}) &= \frac{\Delta \Sigma_{gm}^{\rm res}(R = \chi_l \alpha)}{\Sigma_{crit}} = \frac{2\bar{\rho}_{m}^{\rm 3D}}{\alpha^2 \chi_l^2 \ \Sigma_{crit}} \int_0^{\chi_{\rm lim}} {\rm d} \chi \ \int_0^{R_{\rm min}} {\rm d} R' \ R' \ \xi_{gm}^{\rm res}(\sqrt{R'^2 + \chi^2}) \\
    & \equiv \frac{\mathcal{M}_{t}}{\alpha^2}.
\end{split}
\end{equation}
The last equality defines the so-called \textit{point-mass term} $\mathcal{M}_t$. Its value cannot be worked out analytically with perturbation theory, and so we treat it as a free model parameter. The final result is thus
\begin{equation}
\xi_{t}(\alpha) = \xi_{t}^{\rm PT}(\alpha) + \frac{\mathcal{M}_{t}}{\alpha^2},
\end{equation}
which matches Eq.~\eqref{eq:xi_t} in the main body of the paper. We refer the interested reader to Refs.~\cite{MacCrann2017,Baldauf_2010,Mandelbaum_2010,Pandey_2022,Prat_2022} for more details about point-mass term contributions to the tangential shear 2PCF.

\begin{figure}
\centering 
\includegraphics[width=\textwidth]{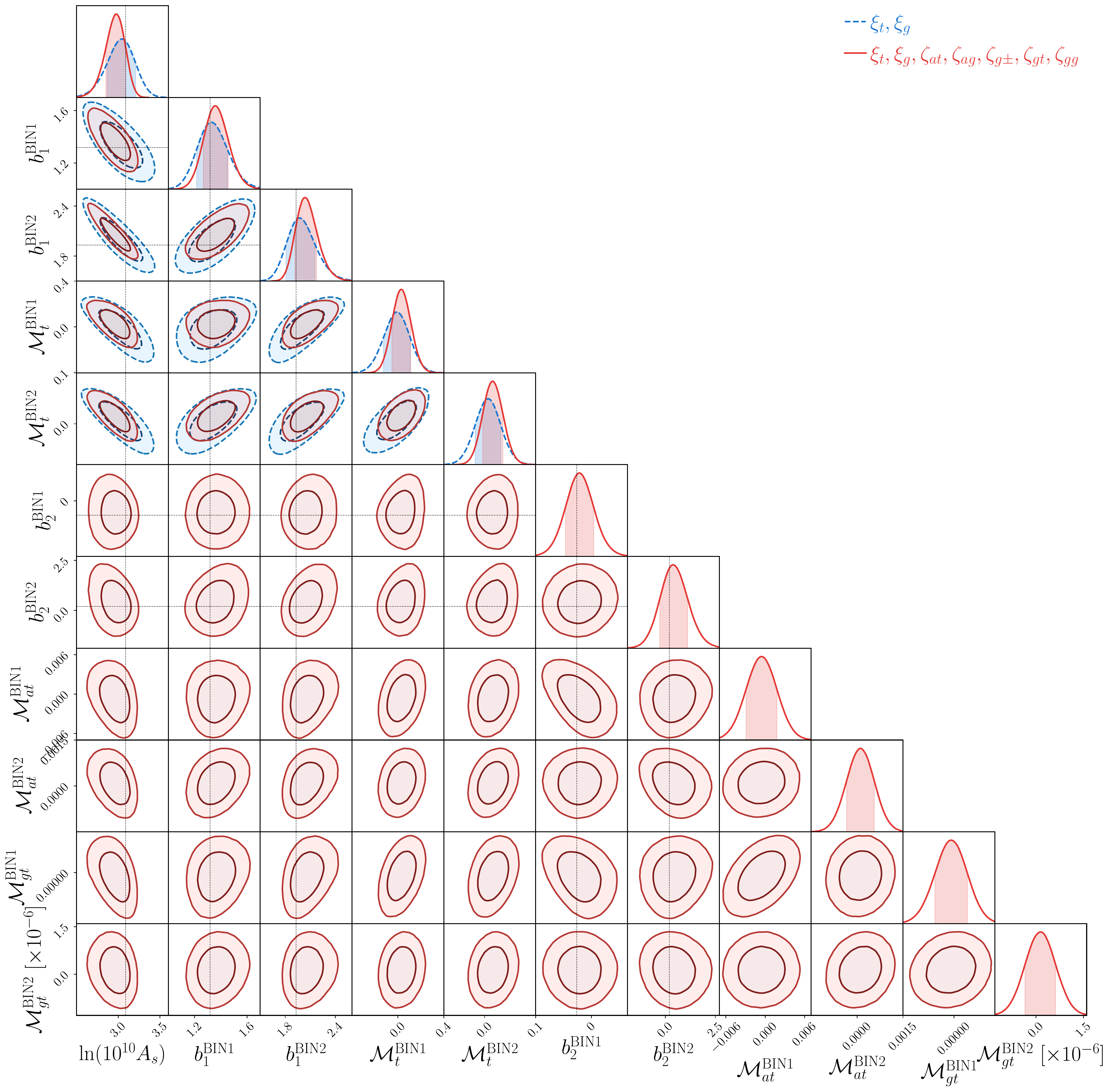}
\caption{\label{fig:MCMC_galaxy_x_shear} Same as Fig.~\ref{fig:MCMC_galaxy_x_shear_marginalised_PM} in the main body of the paper,  but showing also the constraints on the point-mass terms. As there is no analytical way to straightforwardly calculate the expected value of the point-mass terms, these have no dotted lines marking their fiducial values.}
\end{figure}

Similar point mass terms contribute to the integrated 3PCFs as well. Concretely, we can write the \textit{position-dependent tangential shear 2PCF} with a position-dependent point-mass term $\mathcal{M}_{t}(\vtheta_C)$ as
\begin{equation}
    \xi_{t}(\alpha ; \vtheta_C) = \xi_{t}^{\rm PT}(\alpha ; \vtheta_C) + \frac{\mathcal{M}_{t}(\vtheta_C)}{\alpha^2}.
\end{equation}
The correlation of this statistic with the lensing aperture mass $M_{a}(\vtheta_C)$ yields Eq.~\eqref{eq:zeta_at} for $\zeta_{at}(\alpha)$,
\begin{equation}
\label{eq:zeta_at_PM}
\begin{split}
    \zeta_{at}(\alpha) \equiv \langle M_{a}(\vtheta_C) \xi_{t}(\alpha ; \vtheta_C) \rangle &= \langle M_{a}(\vtheta_C) \xi_{t}^{\rm PT}(\alpha ; \vtheta_C) \rangle + \frac{\langle M_{a}(\vtheta_C) \mathcal{M}_{t}(\vtheta_C) \rangle }{\alpha^2} \\
    & = \zeta_{at}^{\rm PT}(\alpha) + \frac{\mathcal{M}_{at}}{\alpha^2},
\end{split}
\end{equation}
where the last equality defines the \textit{point-mass term} for $\zeta_{at}$,  $\mathcal{M}_{at}$. Similarly, the correlation with the mean number of galaxies within apertures yields Eq.~\eqref{eq:zeta_gt} for $\zeta_{gt}(\alpha)$,
\begin{equation}
\label{eq:zeta_gt_PM}
\begin{split}
    \zeta_{gt}(\alpha) \equiv \langle M_{g}(\vtheta_C) \xi_{t}(\alpha ; \vtheta_C) \rangle &= \langle M_{g}(\vtheta_C) \xi_{t}^{\rm PT}(\alpha ; \vtheta_C) \rangle + \frac{\langle M_{g}(\vtheta_C) \mathcal{M}_{t}(\vtheta_C) \rangle }{\alpha^2} \\
    & = \zeta_{gt}^{\rm PT}(\alpha) + \frac{\mathcal{M}_{gt}}{\alpha^2},
\end{split}
\end{equation}
with $\mathcal{M}_{gt}$ a new point-mass term. Again, these new point-mass terms cannot be evaluated with perturbation theory, and so we treat them as free model parameters. 

Figure \ref{fig:MCMC_galaxy_x_shear} shows the constraints from an MCMC analysis with the galaxy 2PCFs (blue) and their combination with the galaxy integrated 3PCFs (red); this is the same as Fig.~\ref{fig:MCMC_galaxy_x_shear_marginalised_PM} in the main body of the paper in Sec.~\ref{sec:validation_MCMC}, except it also shows the constraints on the point-mass terms. Note each of the two lens galaxy samples has their associated point-mass terms. In the figure, it is interesting to note that the addition of the integrated 3PCFs leads to tighter constraints also on the $\mathcal{M}_{t}$ point-mass terms that contribute to $\xi_t$. We leave a more in depth study of the constraints on point-mass terms, including eventual insights on the density profile of the lens galaxies, to future work.

\bibliographystyle{utphys}
\bibliography{references.bib}

\end{document}